\documentclass[%
 aip,
 amsmath,amssymb,
 reprint,%
]{revtex4-1}

\usepackage{graphicx}
\usepackage{dcolumn}
\usepackage{bm}

\usepackage[utf8]{inputenc}
\usepackage[T1]{fontenc}
\usepackage{mathptmx}
\usepackage{etoolbox}
\usepackage{multirow}
\usepackage{makecell}
\usepackage{subfigure}
\usepackage{xcolor}%

\makeatletter
\def\@email#1#2{%
 \endgroup
 \patchcmd{\titleblock@produce}
  {\frontmatter@RRAPformat}
  {\frontmatter@RRAPformat{\produce@RRAP{*#1\href{mailto:#2}{#2}}}\frontmatter@RRAPformat}
  {}{}
}%
\makeatother
\begin{document}

\preprint{AIP/123-QED}

\title[An experimental configuration to study high-enthalpy radiating flows under nonequilibrium de-excitation]{An Experimental Configuration to Study High-Enthalpy Radiating Flows Under Nonequilibrium De-excitation}
\author{Zhuo Liu}
\author{Tiantian Chen}
\author{Jiaao Hao}%
\author{Chih-yung Wen}%
\affiliation{ 
Department of Aeronautical and Aviation Engineering, The Hong Kong Polytechnic University, Kowloon, Hong Kong, People’s Republic of China}%

\author{Qiu Wang}
\altaffiliation{Authors to whom correspondence should be addressed: sangdi.gu@polyu.edu.hk and wangqiu@imech.ac.cn.}
\affiliation{%
State Key Laboratory of High Temperature Gas Dynamics, Institute of Mechanics, Chinese Academy of Sciences, Chinese Academy of Sciences, No.15 Beisihuanxi Road, Beijing 100190, People’s Republic of China}
\author{Sangdi Gu}%
\altaffiliation{Authors to whom correspondence should be addressed: sangdi.gu@polyu.edu.hk and wangqiu@imech.ac.cn.}
\affiliation{ 
Department of Aeronautical and Aviation Engineering, The Hong Kong Polytechnic University, Kowloon, Hong Kong, People’s Republic of China}%
 
\date{\today}

\begin{abstract}
\textcolor{black}{This paper introduces an experimental configuration, called the Prandtl-Meyer plus duct arrangement (PMD), designed to study high-enthalpy radiating flows undergoing nonequilibrium de-excitation. The original design proposed by Wilson, developed without the benefit of modern computational fluid dynamics (CFD), was inadequate for generating a sufficiently large undisturbed zone or achieving a uniform flow along the centerline, necessitating further refinement. Consequently, significant modifications were implemented to enhance PMD's performance, resulting in an expanded undisturbed zone and a uniform centerline flow that facilitate the measurements of nonequilibrium de-excitation.} A general design method is introduced, combining theoretical analysis and numerical simulations to tailor the flow conditions for various research objectives. The implementation involves considerations of the shock tube conditions, PMD configuration, and the effective measurement zone. The interplay between shock tube conditions and airfoil geometry generates diverse de-excitation patterns. The shock tube test time, transition onset location, and radiance intensity determine the effective measurement zone. An example utilizing N\textsubscript{2} as the test gas demonstrates the method, achieving one-dimensional flow with thermal nonequilibrium and chemical freezing along the centerline, validating the method's effectiveness. An effective measurement zone of 200 mm is obtained under this condition, and the primary constraint under high-enthalpy conditions is the limited shock tube test time due to the high shock velocity and low fill pressure.
\end{abstract}

\maketitle

\section{\label{Intro}Introduction}

In recent decades, high-enthalpy radiating flows have attracted considerable attention under the post-shock conditions characterized by internal energy excitation and molecular dissociation \cite{Reynier2016}. However, little attention has been paid to the reverse (expansion) conditions involving internal energy de-excitation and atomic recombination. Computational fluid dynamics (CFD) simulations often employ detailed balance to estimate de-excitation and recombination rates from excitation and dissociation, leading to substantial uncertainties in nonequilibrium modeling for expanding flows. The uncertainties significantly impact the design of the afterbody of atmospheric entry vehicles, where radiative heating, influenced heavily by vibrational and electronic de-excitation, plays a dominant role \cite{Surzhikov2007}. Furthermore, hypersonic impulse facilities, such as shock tunnels and expansion tunnels, also suffer from the same uncertainties under high-enthalpy conditions, resulting in uncertainties in interpreting experimental data \cite{Ray2020,MacLean2006}.

Past studies on expanding flows predominantly focused on low-enthalpy conditions, where electronic excitation and chemical reactions were often absent. These experiments revealed that the vibrational relaxation time, $\tau_\mathrm{v}$, under de-excitation conditions for N\textsubscript{2}, CO, and CO\textsubscript{2} was around 5-70\cite{Sharma1993}, 1-1000 \cite{McLaren1970,Russo1967}, and 1.06-1.14\cite{Ibraguimova2012} times shorter, respectively, than their corresponding values under excitation conditions at the same pressure and temperature. Theoretical explanations were inadequate to account for this discrepancy. While vibrational de-excitation has been studied, electronic de-excitation and atomic recombination under high-enthalpy conditions remain under-explored. Due to the limitations of traditional experimental techniques, vibrational temperature measurements were primarily performed using the line or band reversal method, which involves simultaneous measurements of emission and absorption\cite{Blom1968}. Modern techniques like tunable diode laser absorption spectroscopy (TDLAS) and optical emission spectroscopy (OES) offer better alternatives for measuring the gas properties and nonequilibrium radiation in hypersonic impulse facilities. Therefore, more experiments under higher-enthalpy conditions, utilizing advanced spectroscopy, are necessary to achieve a deeper understanding of de-excitation processes and to enhance numerical models.

Various numerical and theoretical studies on de-excitation and recombination have been conducted in diverse scenarios, including hypervelocity capsule entry (or reentry), nozzle expansion, Prandtl-Meyer expansion, etc. Characterizing expanding flow is crucial for the design of the afterbody thermal protection system. For Earth reentry, Mars entry, and Titan entry, the radiative heating typically exceeds the convective heating in the afterbody region \cite{Brandis2020,Johnston2019,Johnston2015}, showing the importance of nonequilibrium radiation. The quasi-one-dimensional nature of hypersonic nozzle flows allows for the application of advanced models such as the multi-temperature model\cite{Park1995}, the state-to-state model\cite{Gu2022}, or the Monte Carlo approach\cite{Gimelshein2021}. These sophisticated models often reveal distinct vibrational and electronic temperatures for different species, highlighting the complexity of the internal energy distribution.

The study of expanding flows began experimentally in the 1960s and 1970s, focusing mainly on vibrational de-excitation under low-enthalpy conditions. To observe de-excitation or recombination, the gas must first be heated and then cooled to a lower temperature. Heating is typically achieved using a shock tube or electric arc, while cooling can be done by mounting an acceleration tube filled with low-pressure gas or a nozzle in the end wall. These methods correspond to unsteady expansion and nozzle expansion, respectively. Both configurations usually generate a high degree of expansion and result in essentially one-dimensional or quasi-one-dimensional flow. Nevertheless, unsteady expansion lacks spatial resolution, making it unsuitable for diagnosing radiating flow using OES. A frozen-equilibrium profile can be obtained by measuring the gas properties at the exit of the acceleration tube. Holbeche\cite{Holbeche1964} reported the spectrum-line-reversal measurements in an expansion tube with O\textsubscript{2} as the test gas and found that the measured temperature was higher than the theoretical calculations. Cleaver and Crow\cite{Cleaver1973} investigated the vibrational relaxation of oxygen and concluded that the de-excitation rate closely matches that under the excitation conditions. Nasser and Cleaver\cite{Nasser1977} showed that vibrational de-excitation is twice as fast as the excitation condition for CO, while Mclaren and Appleton\cite{McLaren1970} found the relaxation rates to be the same in both compression and expansion environments.

Two different configurations can be used for nozzle expansion, distinguished by the presence or absence of a diaphragm separating the shock tube from the nozzle. Without a diaphragm, a diverging nozzle is required to expand the shock-heated gas\cite{Tibère-Inglesse2022}. In this configuration, the test time is essentially the same as the shock-tube test time, and the flow remains unsteady, with the nozzle inflow. However, generating a pronounced expansion is difficult due to the relatively low inlet-to-outlet geometry ratio. A low degree of expansion allows for the measurement of nonequilibrium radiation. Recently, Tibère-Inglesse et al. mounted an expansion cone to the end of the EAST shock tube to investigate the radiance emitted from expanding air and found that CFD underpredicts the measured radiance \cite{Tibère-Inglesse2022}. When a diaphragm is present at the end wall, the incident shock reflects off the shock tube end wall, generating a high-temperature reservoir. In this case, a converging-diverging nozzle is used to achieve steady expansion. Since the throat radius can be considerably small, the reservoir gas takes longer to flow out, resulting in a significantly longer test time compared to unsteady expansion, typically around several milliseconds. This configuration faces uncertainties in characterizing the high-enthalpy reservoir gas, and the large degree of expansion leads to weak radiative emission. Nonequilibrium expansion of N\textsubscript{2}\cite{Sharma1993,Sharma1993VR,Sebacher1974} and CO\cite{Russo1967,Bender1978} has been extensively investigated using this configuration, showing faster relaxation rates in expanding flow compared to excitation conditions. Additionally, the nozzle can also serve as a test model when positioned in the test section of a ground facility. Park utilized this method to measure the ionic recombination rate of N\textsubscript{2}\cite{Park1968}, as well as the electronic excitation temperature and electron temperature in a plasma-jet wind tunnel\cite{Park1973}.

Additionally, cooling can also be achieved through Prandtl-Meyer expansion using a two-dimensional test model. The first method to generate the Prandtl-Meyer expansion involves directing the gas flow over a corner. Depending on whether the freestream is parallel to the pre-corner portion of the model, the test models can be referred to as \textit{wedge model} or \textit{wall model}\cite{Igra1975}. The wall model is typically positioned at the nozzle outlet, allowing for a significantly larger expansion angle than the wedge model, or it can be adjusted to create varying degrees of expansion. An appropriate expansion angle makes it possible to measure the nonequilibrium radiation above the model after expansion. However, this model produces a thick boundary layer, and the gas undergoes additional shock heating before expansion, introducing more uncertainties in the pre-expansion gas.

Recently, investigations of thermochemical nonequilibrium with CO\textsubscript{2}\cite{Gu2021}, Ar\cite{Kelly2021}, and air\cite{Wei2017} as test gases were conducted using the wall model in the X2 expansion tube. Measurements were typically taken several millimeters above the test model. A faster relaxation rate was observed for CO\textsubscript{2} in expanding flow, and significant discrepancies exist when comparing the measured radiance with the simulation result for air. Electromagnetic radiation from upstream was observed for Ar, suggesting a complex interaction between the test gas and the radiation.

The wedge model is typically positioned at the shock tube outlet, where the freestream is well determined. This model creates a thin boundary layer, with the expansion angle kept small enough to prevent detachment of the shock at the leading edge. Using this model, Igra\cite{Igra1975} measured the three-body electron-ion-electron collisional recombination rate of Ar. Another version of the wedge model consists of two airfoilss (wedges) positioned symmetrically in the vertical direction, as depicted in Fig.~\ref{fig:PMandST}. This configuration allows the gas to flow over two symmetric corners, thereby generating double Prandtl-Meyer expansion fans \cite{Wilson1963}. The flow on the centerline can be treated as one-dimensional and decoupled from the fluid flow. To exclude external disturbances, a duct can be placed on the centerline, facilitating the measurements of de-excitation and recombination. Since two wedge models are essentially employed in this configuration, it is typically positioned at the shock tube outlet, and the main limitation also comes with the expansion angle and relatively low degree of expansion compared to other methods. This configuration, called the Prandtl-Meyer plus duct arrangement (PMD), was first used by Wilson to measure the oxygen recombination rate\cite{Wilson1963}, and reasonable agreement was observed compared with the dissociation rate. Blom and Pratt\cite{Blom1969} measured the relaxation rate of CO both behind the incident shock and downstream of the expansion, observing a higher relaxation rate in the expanding flow.

To study the high-enthalpy radiating flow, among these experimental configurations, the wedge model, wall model, and diverging nozzle all have a spatial resolution, facilitating the OES measurements and generating a moderate or low degree of expansion, resulting in sufficient radiance after expansion. PMD offers a higher degree of expansion than the ordinary wedge model and diverging nozzle. Furthermore, the centerline flow of PMD is steady and can be treated one-dimensionally, facilitating the evaluation of numerical models. Compared with nozzle expansion or unsteady expansion, PMD is rarely utilized and requires a lower manufacturing cost. The adjustment of the expansion degree in PMD is highly flexible, showing strong adaptability to different gases and research objectives. Recently, Gu et al.\cite{Gu2024} highlighted an important application of PMD under non-radiating conditions: the experimental determination of the equilibrium dissociation/recombination rates. The overall atomic recombination process is uninfluenced by thermal nonequilibrium for binary mixtures of O\textsubscript{2}/O and N\textsubscript{2}/N when the translational temperature is lower than around 500~K, and the double Prandtl-Meyer expansion creates a rapid decrease in temperature, making it well-suited for equilibrium rate measurements. Therefore, there is an significant interest in using PMD to study nonequilibrium expanding flows in general.

In this paper, we utilize and investigate PMD to study the high-enthalpy radiating flow. Based on Wilson's design, important modifications were made to enhance the performance. A general method for designing PMD is also presented, based on theoretical analysis and numerical simulations, applicable to any nonequilibrium flow (radiating or not). Section~\ref{Metho} elaborates on the experimental method, including PMD, spectroscopy diagnostics, numerical methods, mesh independence study, and Mirels’ theory\cite{Mirels1963} for determining the effective test time. airfoil design is the focus in Section~\ref{QualiStudy}, with qualitative aspects investigated and several modifications implemented. The design method summarized in a flowchart will be presented in Section~\ref{DesignoP}. In Section~\ref{AExample}, we provide an example of high-enthalpy radiating N\textsubscript{2} flows using the design method. Finally, the main conclusions will be presented in Section~\ref{Conclusions}.

\section{\label{Metho}Methodology}
\subsection{\label{PMEA}the Prandtl-Meyer plus duct arrangement}
\begin{figure*}
 \centering
	\includegraphics[width=0.9\textwidth]{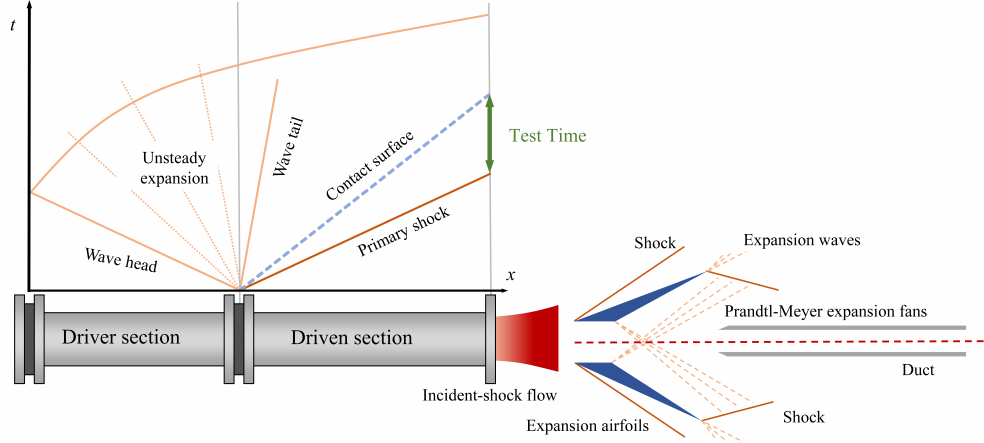}
	\caption{\label{fig:PMandST}A Two-dimensional Prandtl-Meyer plus duct arrangement was placed at the exit of a conventional shock tube. The red line indicates the location of the emission spectroscopy measurement.}
\end{figure*}

Figure~\ref{fig:PMandST} shows the original design of PMD by Wilson\cite{Wilson1963}. The flow undergoes rapid expansion upon passing through the double Prandtl-Meyer expansion fans, resulting in a sharp decrease in the translational-rotational temperature while chemical, vibrational, and electronic modes remain frozen. Subsequently, the flow enters a constant-area duct where vibrational and electronic de-excitation can occur, along with the atomic recombination. This configuration creates opposite states of the flow behind a normal shock wave.
The Prandtl-Meyer expansion experiment is typically conducted in a shock tube to avoid undesirable and unnecessary processes in wind tunnels. Therefore, the freestream is well-determined, and the spanwise length can be the diameter of the shock tube minus the boundary layer thickness. The flow process of a shock tube is illustrated in Fig.~\ref{fig:PMandST}. When implementing this configuration, three aspects require consideration: the shock tube conditions, PMD configuration, and the effective measurement zone. Wilson\cite{Wilson1963} and Slack\cite{Slack1969} designed this configuration primarily through a theoretical analysis of flow conditions, focusing on specific design aspects. Wilson determined the shock tube conditions and expansion angles through one-dimensional calculations. Meanwhile, Slack analyzed the start-up process, airfoil geometry, and gas state during expansion and emphasized the importance of achieving equilibrium behind the incident shock. \textcolor{black}{However, in the absence of modern CFD , previous researches on PMD lack a detailed geometric design for the airfoil and a comprehensive investigation of the flow field. Consequently, the original design proposed by Wilson fails to generate a sufficiently large undisturbed zone and a uniform centerline flow, as shown in Fig.~\ref{fig:DegComp} and \ref{fig:Temperaturepa}, making it inadequate for radiation measurements. The modifications to PMD will be elaborated upon in subsequent chapters. A more comprehensive design approach can be realized using modern CFD.} The simulations of the configuration in this paper are currently two-dimensional, but three-dimensional effects cannot be ignored for an actual model. To prevent the expansion in the spanwise direction, optical windows should be installed on each side of the airfoils and the duct, although this may introduce leading-edge shocks and boundary layers due to viscous interactions.

\subsection{\label{OES}Optical emission spectroscopy}
The flow inside the duct along the centerline, indicated by the red dashed line in Fig.~\ref{fig:PMandST}, exhibits uniform characteristics and can be regarded as one-dimensional. The calibrated OES data provides one-dimensional spatial information, making it well-suited for diagnosing the high-enthalpy radiating gas using this configuration. Along the centerline, the entire relaxation can be measured simultaneously. To simulate the radiance originating from the high-enthalpy gas, absorption and emission coefficients are initially calculated by SPARK (Simulation Platform for Aerodynamics Radiation and Kinetics)\cite{daSilva2021}, which will be briefly introduced in section \ref{SPARK}. Accounting for the concurrent effects of emission and absorption of a gas mixture, the radiance can be expressed by\cite{Anderson2006} 
\begin{equation}\label{eq:dIds}
\frac{\mathrm{d}I_{\lambda}}{\mathrm{d}s}=J_{\nu}-k_{\nu}I_{\nu}
\end{equation}
where $I_\mathrm{\lambda}$ represents the spectral radiance at a specific wavelength, $J_\mathrm{\nu}$ and $I_\mathrm{\nu}$ are the emission and absorption coefficients, respectively, and ds is the distance element transmitted in a mixture. By integrating this equation from $x = 0$, and assuming constant values for $J_\mathrm{\nu}$ and $I_\mathrm{\nu}$, we obtain the spectral radiance crossing uniform matter: 
\begin{equation}\label{eq:I}
	I_{\lambda, s} = \frac{J_{\nu}}{k_{\nu}}(1 - e^{-k_{\nu}s}) + I_{\lambda, 0}e^{-k_{\nu}s}
\end{equation}
In this study, the initial radiance $I_\mathrm{\lambda,0}$ is set to 0, and the transmitted distance is set to the diameter of the shock tube. The radiance intensity must be strong enough to facilitate the camera observation and achieve high signal-to-noise ratio data. Hence, the threshold of spectral radiance is set to 0.1 W/cm\textsuperscript{2}/$\mu$m /sr to satisfy this requirement.

\subsection{\label{NM}Numerical methods}
\subsubsection{\label{Eilmer4}Eilmer4}
Simulations were performed using a thermochemical nonequilibrium model with the open-source code Eilmer4\cite{Gibbons2023}. Eilmer4 is a finite-volume-based code specifically designed for transient, compressible flow simulations in two and three spatial dimensions. The equilibrium shock solver SDToolbox\cite{Browne2018} was used to calculate the freestream conditions for Eilmer4 by inputting the shock speed and initial pressure. The thermochemical nonequilibrium phenomenon was accounted for by employing Park's two-temperature model, which considers the translational-rotational temperature (\textit{T}\textsubscript{tr}) and vibrational-electronic-electron temperature (\textit{T}\textsubscript{vee}). A chemical model consisting of five species (N\textsubscript{2}, N, N\textsubscript{2}\textsuperscript{+}, N\textsuperscript{+}, e\textsuperscript{-}) was employed to account for chemical nonequilibrium\cite{Kim2021}.

\subsubsection{\label{SPARK}SPARK}
SPARK is a Line-by-Line code for simulating high-temperature, low-pressure plasma radiation, with an emphasis on radiation from entry plasmas, plasma sources, and high-temperature plasma sources radiation\cite{daSilva2021}. The simulations require input parameters such as translational-rotational, vibrational, and electronic temperature, as well as particle number density. In this study, a two-temperature model is employed, where the vibrational, electronic temperatures are assumed to be equal when calculating the emission and absorption coefficients. Subsequently, radiance from the high-enthalpy gas can be computed using Eq.~(\ref{eq:dIds}). When simulating the experimentally measured spectra, a Gaussian function with a full width at half maximum (FWHM) of 0.2 nm is applied to account for apparatus broadening.

\subsection{\label{MIS}Mesh independence study}
\begin{figure}
	\subfigure[Domain A]{\label{fig:GridsA}\includegraphics[width=0.45\textwidth]{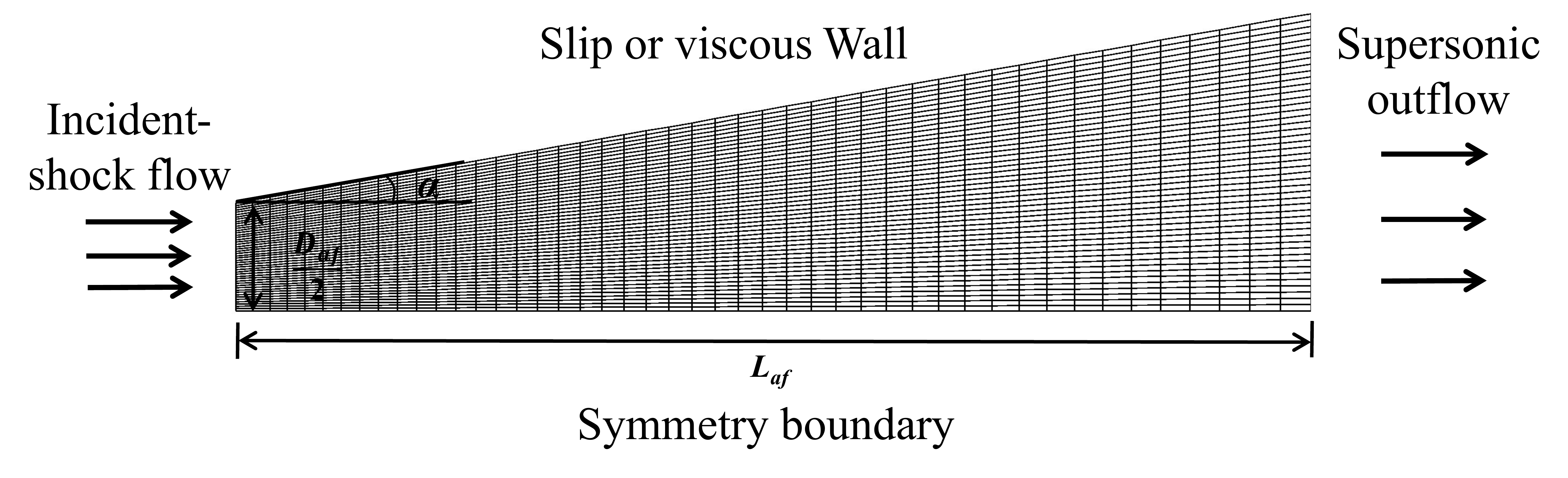}}
	\subfigure[Domain B]{\label{fig:GridsB}\includegraphics[width=0.45\textwidth]{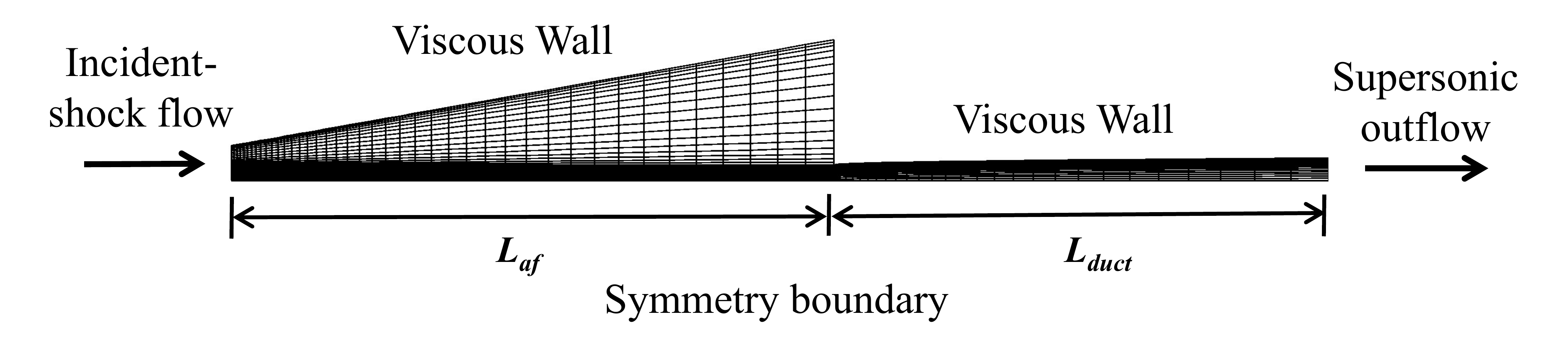}}
	\caption{\label{fig:Grids}Computational domains for (a) the determination of airfoil geometry and (b) simulating the flow-establishing process}
\end{figure}

PMD design was conducted with two different domains, as shown in Fig.~\ref{fig:Grids}. Domain A is employed to determine the shock tube conditions and airfoil geometry. Domain B is used to simulate the flow establishment process. The freestream conditions are listed in Table~\ref{tab:shockWC}, and condition 4 is used for the mesh independence study.

\begin{figure}
	\subfigure[Domain A, temperature]
	{\label{fig:GIS-GA-T}\includegraphics[width=0.45\textwidth]{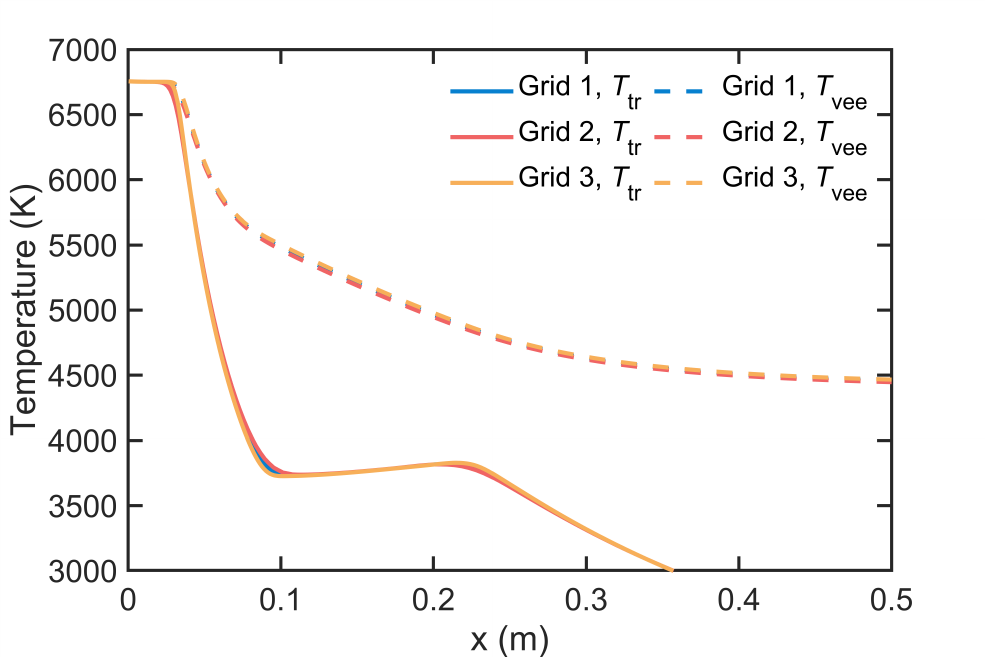}}
	\subfigure[Domain A, pressure]
	{\label{fig:GIS-GA-P}\includegraphics[width=0.45\textwidth]{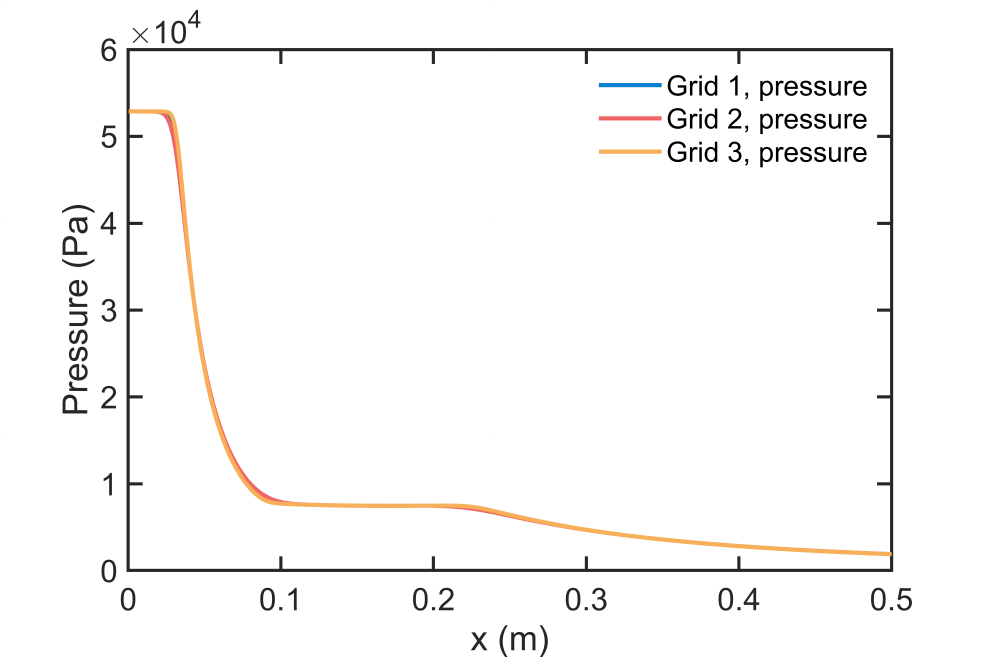}}
	\subfigure[Domain B, temperature]
	{\label{fig:GIS-GB-T}\includegraphics[width=0.45\textwidth]{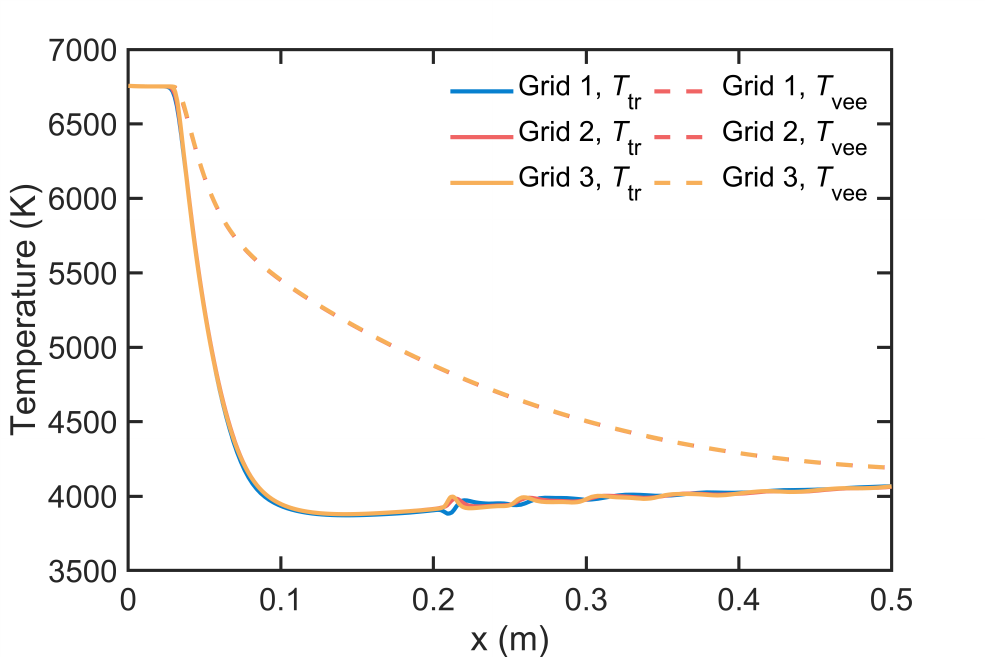}}
	\subfigure[Domain B, pressure]
	{\label{fig:GIS-GB-P}\includegraphics[width=0.45\textwidth]{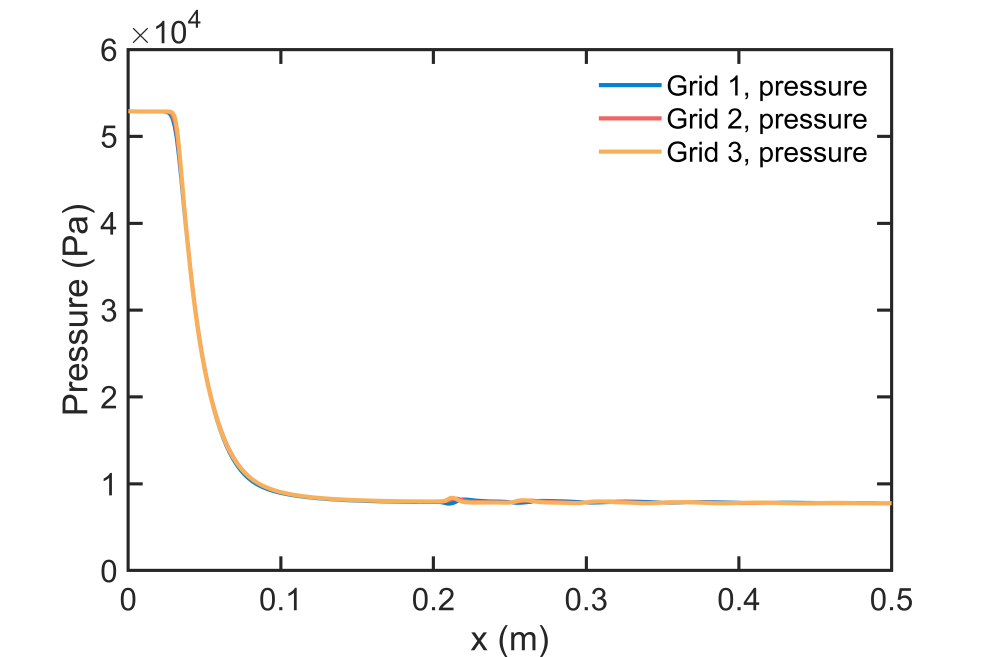}}
	\caption{\label{fig:GIS}Comparison of the (a) nonequilibrium temperatures and (b) pressure distributions along the centerline under domain A and B with different cell numbers.}
\end{figure}

Figure \ref{fig:GIS}(a,b) compares the temperature and pressure distributions using domain A. The geometry of the computational domain is $\alpha=10^{\circ}$, \textit{D}\textsubscript{af} = 22 mm, and \textit{L}\textsubscript{af} = 0.5 m. The number of cells in rids 1, 2, and 3 are 4.8 k ($\mathrm{streamwise\times wall-normal}$: $140\times34$), 9.6 k ($200\times48$), and 22 k ($300\times72$), respectively. Due to the small influence of the viscous effect on the expansion, an inviscid boundary condition can be utilized for domain A. Only minor discrepancies were observed at the start and end of expansion fans for different grids. Therefore, grid 1 was chosen for further simulations.

When the duct was positioned downstream of the airfoils, the simulations were performed using domain B, with $L_\mathrm{af}$ = 0.19 m and $L_\mathrm{duct}$ = 0.4 m, as shown in Fig.~\ref{fig:GIS}(c,d). The numbers of cells for grids 1, 2, and 3 are 13 k ($\mathrm{streamwise\times wall-normal}$: $100\times84+100\times42$), 26 k ($144\times120+144\times60$), and 58 k ($216\times180+216\times90$), respectively. Viscosity must be incorporated in the simulations, as it influences the flow establishment rate. The distributions among the three grids are almost identical before \textit{x} = 0.19 m, with minor discrepancies observed at the duct front, where a small disturbance is generated. The duct geometry in domain B was modified based on the displacement thickness (disussed in Section~\ref{DuctSO}). However, this adjustment is unlikely to be perfectly tailored to the inlet flow, leading to weak expansion or compression waves at the duct's leading edge. For grid 3, a slight temperature dip is initially observed at \textit{x} = 0.19 m, but this anomaly diminishes with increasing grid resolution. However, this small disturbance is well characterized by grid 2 and grid 3, and the disturbance generated in these two grids is almost the same. Therefore, grid 2 was selected for further simulations.
%
\subsection{\label{Mirels}Mirels’ theory}
Mirels proposed a theoretical method to evaluate shock tube test time\cite{Mirels1963}. For an ideal shock tube, the contact surface and the shock both move downstream at constant velocities, and the flow between them is uniform. However, in an actual shock tube, the existence of the boundary layer can cause the contact surface to accelerate and the shock to decelerate until both reach the same velocity, and the flow between them is nonuniform. Mirels established a shock-stationary coordinate system to make the flow between the contact surface and the shock steady. His study provides a relationship between the shock tube length and the effective test time, as well as the maximum separation distance between the shock and the contact surface and the corresponding maximum test time. This paper only considers the laminar wall boundary layer effect and detailed theoretical derivations can be found in Ref.~\onlinecite{Mirels1963}.
\section{\label{QualiStudy}Qualitative Study of the Flow Field Around the Airfoil}
\subsection{\label{UndisZone}The undisturbed zone}
According to Wilson’s design\cite{Mirels1963}, PMD comprises two airfoils and a constant area duct, as shown in Fig.~\ref{fig:PMandST}. Prandtl-Meyer expansion waves are generated by the airfoils to create opposite states of the flow behind a normal shock wave. The duct captures the central flow and excludes external disturbances that may intrude on the centerline flow.

The feasibility of PMD was first investigated with only the airfoil placed at the shock tube exit. In a supersonic flow environment, PMD ensures that the gas entering the duct is solely influenced by the airfoil. The primary waves generated around the airfoil are depicted in Fig.~\ref{fig:WisDeg}. At the leading edge of the airfoil, two shocks form on both sides: an oblique shock on the upper side due to the inclined plane, and a leading-edge shock on the lower side due to viscous interaction. Expansion waves and oblique shocks are generated to balance the velocity, pressure, and temperature of the two flows at the tail of the airfoil. The shock, propagating in the lower right direction, will significantly affect the centerline flow, resulting in the interruption of the relaxation process after a short distance. The short distance labeled as \textit{D}\textsubscript{4} is indicated by the green line in Fig.~\ref{fig:WisDeg}. Because no disturbance occurs on \textit{D}\textsubscript{4}, it can be called the undisturbed zone. 

Three parameters define the geometry of the airfoil in Wilson’s design: the distance between two airfoils (\textit{D}\textsubscript{af}), the expansion angle ($\alpha$), and the horizontal length of the expansion section (\textit{L}\textsubscript{af}). The gas properties after crossing the Prandtl-Meyer expansion waves are primarily determined by $\alpha$. All three parameters can influence the size of the undisturbed zone (\textit{D}\textsubscript{4}); since \textit{D}\textsubscript{4} is determined by the distance between the tail of Prandtl-Meyer expansion waves and the location of the curved shock on the centerline. For simplicity, assuming that all the trajectories of expansions and curved shock propagate along straight lines, the length of the undisturbed zone can be expressed as follows: 

\begin{figure}
	\subfigure[Wilson’s airfoil]
	{\label{fig:WisDeg}\includegraphics[width=0.45\textwidth]{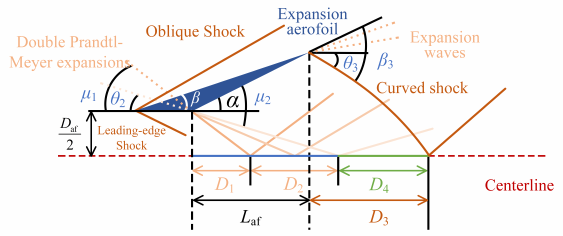}}
	\subfigure[Modified airfoil]
	{\label{fig:ModiDeg}\includegraphics[width=0.45\textwidth]{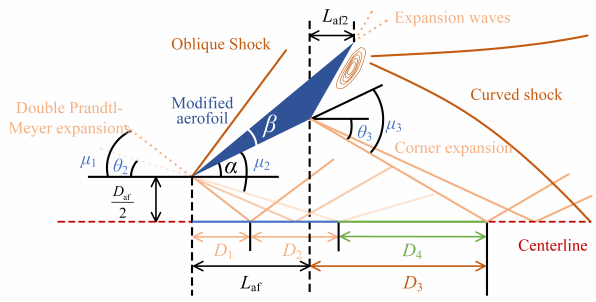}}
	\caption{\label{fig:airfoils}Primary waves around the airfoil in Wilson’s and modified design.}
\end{figure}
\begin{eqnarray}\label{eq:UndisZone}
	D_{4} &=& L_\mathrm{af} + D_3 - D_1 - D_2\nonumber \\
	      &=&L_\mathrm{af}(1 + \tan \alpha \cot \theta_3) +\frac{D_\mathrm{af}}{2}(\cot \theta_3 - \cot \theta_2)
\end{eqnarray}

\subsection{\label{ModitoWil}Modifications to Wilson’s airfoil}
In a supersonic flow field, Fig.~\ref{fig:WisDeg} shows that the variation of $\theta_\mathrm{3}$ is independent of $D_\mathrm{af}$, $L_\mathrm{af}$, and $\alpha$, and $D_\mathrm{4}$ increases as $\theta_\mathrm{3}$ decreases. $\theta_\mathrm{3}$ can be reduced by replacing the curved shock wave with expansion waves. Furthermore, the pre-corner portion should be removed as it generates the leading-edge shock. The modified airfoil and its wave diagram are shown in Fig.~\ref{fig:ModiDeg}.

In the modified design, assuming that all the trajectories of expansion waves propagate along straight lines again. For the expansion angle, the variation in $\alpha$ will concurrently influence $\theta_\mathrm{2}$ and $\theta_\mathrm{3}$. From Fig.~\ref{fig:ModiDeg}, we can infer $\theta_\mathrm{2}=\mu_\mathrm{2}-\alpha$, $\theta_\mathrm{3}=\mu_\mathrm{3}-\alpha$, with $\mu_\mathrm{1}$, $\mu_\mathrm{2}$, and $\mu_\mathrm{3}$ being Mach angles, determined only by the local Mach number. When $\alpha$ increases, the flow expands into a higher Mach number, leading to a decrease in $\mu_\mathrm{1}$ and $\mu_\mathrm{2}$. Due to the proximity of the Prandtl-Meyer expansion waves to the tail of the airfoil, the flow around the tail of the upper airfoil will cross more expansion waves, which increases the Mach number around the corner of the airfoil, resulting in $\theta_\mathrm{3}-\theta_\mathrm{2}<0$ and $\cot\theta_\mathrm{3}-\cot\theta_\mathrm{2}>0$. Although determining the variation of $\cot\theta_\mathrm{3}-\cot\theta_\mathrm{2}$ is difficult, $L_\mathrm{af}$ is consistently much larger than $D_\mathrm{af}/2$ in most cases, making the first term in Eq.~\ref{eq:UndisZone} more influential. Therefore, the undisturbed zone enlarges as $\alpha$ increases.

For $D_\mathrm{af}$, the change is independent of $\theta_\mathrm{2}$ and $\theta_\mathrm{3}$, and $\cot\theta_\mathrm{3}-\cot\theta_\mathrm{2}$ is greater than 0 from the above analysis. Consequently, the undisturbed zone will always enlarge when $D_\mathrm{af}$ increases. Similarly, the change of $L_\mathrm{af}$ is also independent of $\theta_\mathrm{2}$ and $\theta_\mathrm{3}$, and will only affect the location of the corner expansion on the centerline. Therefore, the undisturbed zone will also enlarge when $L_\mathrm{af}$ increases. The influence of the three geometric parameters on the flow field is summarized in Table~\ref{tab:InfluGeo}.
\begin{table*}
 \renewcommand{\arraystretch}{1.2}
	\caption{\label{tab:InfluGeo}The influence of the geometric parameters of the airfoil on the flow field structure.}
	\begin{ruledtabular}
		\begin{tabular}{lccccc}
			Parameters 
			&Effect on $\theta_\mathrm{2}=\mu_\mathrm{2}-\alpha$ 
			&Effect on $\theta_\mathrm{3}=\mu_\mathrm{3}-\alpha$
			&Effect on $D_\mathrm{1}-D_\mathrm{2}$
			&Effect on $D_\mathrm{3}$ &Effect on $D_\mathrm{4}$\\
			\hline
			$\alpha\uparrow$        &$\downarrow$ &$\downarrow$ &$\uparrow$ &$\uparrow$ &$\uparrow$ \\
			$D_\mathrm{af}\uparrow$ &Unchanged    &unchanged    &$\uparrow$ &$\uparrow$ &$\uparrow$ \\
			$L_\mathrm{af}\uparrow$ &Unchanged    &Unchanged    &Unchanged &$\uparrow$ &$\uparrow$ \\
		\end{tabular}
	\end{ruledtabular}
\end{table*}

\begin{figure}
	\centering
	\includegraphics[width=0.45\textwidth]{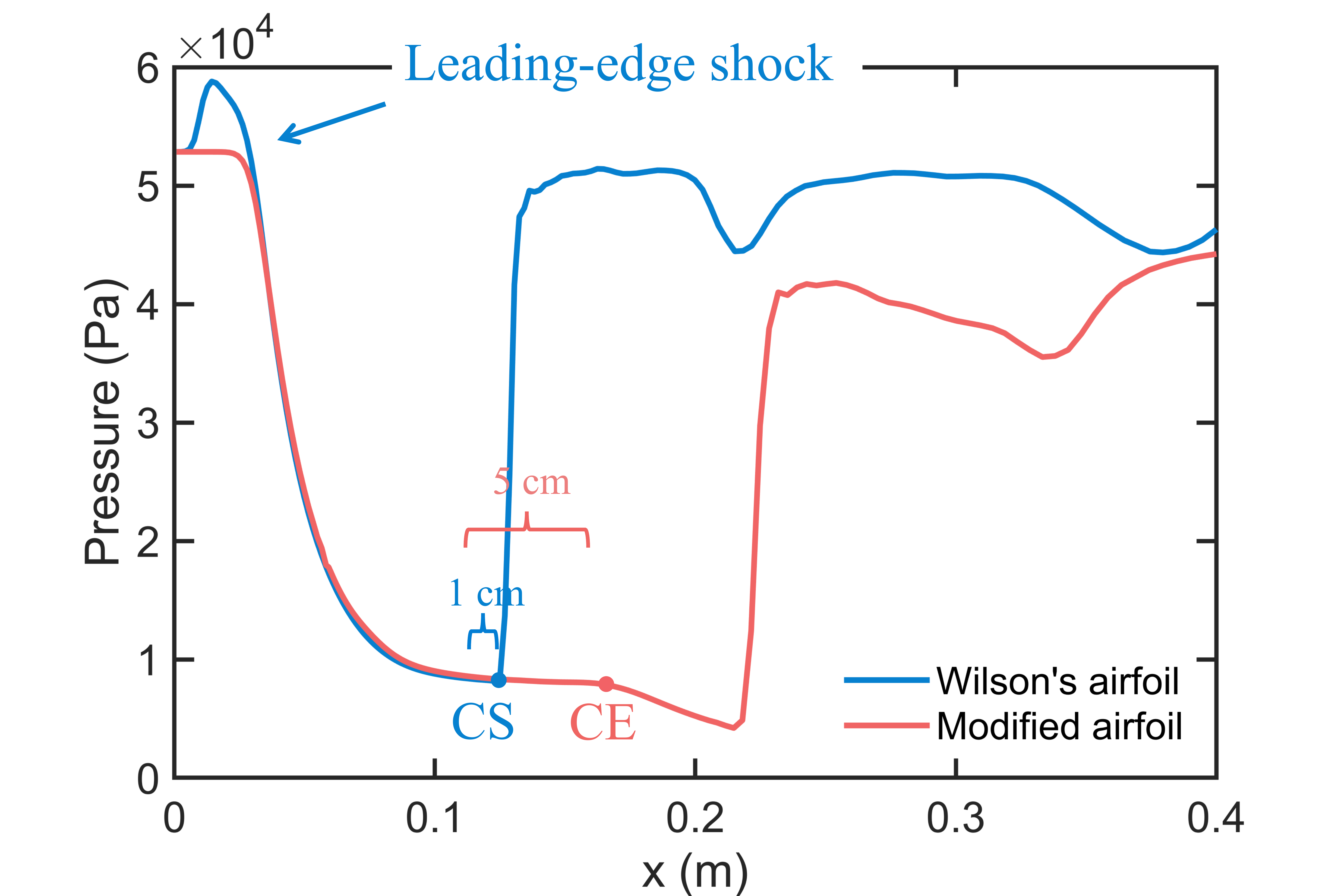}
	\caption{\label{fig:DegComp}Pressure distribution along the centerline with different airfoils. CS or CE indicates the undisturbed zone is interrupted by the curved shock wave or the corner expansion waves.}
\end{figure}

Simulations of the two airfoils were performed to verify the influence of the modifications, and condition 4 in Table~\ref{tab:shockWC} is used as the freestream. When different airfoils were employed, the pressure distribution along the centerline is compared in Fig.~\ref{fig:DegComp}. The parameters of the two airfoils were set to the same value: $\alpha=10^{\circ}$, $D_\mathrm{af}=22$ mm, and $L_\mathrm{af}=60$ mm. There are two more parameters to define the geometry of the modified airfoil ($L_\mathrm{af2}$ and $\beta$), which are set to $L_\mathrm{af2}=30$ mm and $\beta=10^{\circ}$. When Wilson’s airfoil is employed, a leading-edge shock is formed at the pre-corner portion, and the undisturbed zone is interrupted by the corner shock, resulting in the undisturbed zone being too short to be measured. In contrast, the undisturbed zone becomes much larger when the modified airfoil is employed because the shock angle $\theta_\mathrm{3}$ in Wilson’s design is larger than the expansion head angle $\theta_\mathrm{3}$ in the modified design. 

\begin{figure*}
	\subfigure[]{\label{fig:CharaCompBef}\includegraphics[width=0.45\textwidth]{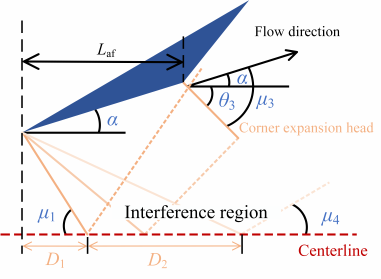}}
	\subfigure[]{\label{fig:CharaCompAft}\includegraphics[width=0.45\textwidth]{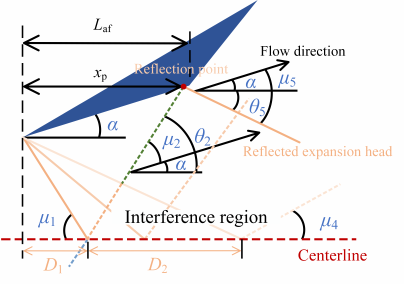}}
	\caption{\label{fig:CharaComp}Wave diagrams just (a) before and (b) after the Prandtl-Meyer expansion wave reflected on the lower wall of the airfoil. Solid and Dashed lines represent type $C_{-}$ and $C_{+}$ characteristic lines, respectively.}
\end{figure*}

According to the above analysis, $L_\mathrm{af}$ does not affect $\theta_\mathrm{2}$ and $\theta_\mathrm{3}$, and increasing $L_\mathrm{af}$ can continuously expand the undisturbed zone. However, this approach becomes impractical when considering the reflections of the Prandtl-Meyer expansion waves. In Fig.~\ref{fig:CharaComp}, the wave diagram in the double Prandtl-Meyer region is illustrated, with the reflection point marked as a red dot. If the length of the airfoil exceeds a certain threshold, the Prandtl-Meyer expansion waves will reflect off the lower wall of the airfoil. Conversely, a short airfoil causes premature disruption of the centerline flow by the corner expansion.

Reflection becomes possible when $L_\mathrm{af}$ reaches a critical length. The method of characteristics is required to determine if the reflected expansion reaches the centerline before the corner expansion. However, this evaluation can be simplified using a scenario depicted in Fig.~\ref{fig:CharaComp}, which shows the wave diagrams just before and after the Prandtl-Meyer expansion wave reflected on the lower wall of the airfoil. Both the reflected expansion and corner expansion are $C_{-}$ characteristic lines\cite{Anderson2011}, with horizontal angles $\theta=\mu-\alpha$. Since the flow directions crossing both expansion heads are the same, the location at which the expansion waves reach the centerline depends solely on the Mach angles. The reflected expansion head crosses part of the interference region composed of the Prandtl-Meyer expansion wave and the reflected expansion, as shown in Fig.~\ref{fig:CharaCompBef}. Consequently, a higher Mach number is presented at the root of the reflected expansion head, resulting in a smaller Mach angle and thereby a delayed arrival on the centerline compared to the corner expansion. Therefore, the size of the undisturbed zone is limited by the reflection of the Prandtl-Meyer expansion. To maximize the undisturbed zone, $L_\mathrm{af}$ must be larger than \textit{x}\textsubscript{p} in both Wilson’s and the modified design. However, the modified airfoil significantly reduces the required length since $\theta_\mathrm{3}$ is larger than that of Wilson’s airfoil.

The two added geometric parameters ($L_\mathrm{af}$, $\beta$) do not influence the double Prandtl-Meyer expansions; thus, the undisturbed zone is still primarily determined by Eq.~\ref{eq:UndisZone}. Although the influence of the leading-edge shock on the gas properties behind the expansion is small under condition 4, the viscosity becomes more influential as Mach number increases. Compared to Wilson’s airfoil, which requires distinct airfoils for different expansion angles, the modified airfoil simplifies the model by enabling a single airfoil to achieve various expansion angles through rotation. Nevertheless, the larger upper-side deflection angle restricts the maximum expansion angle, leading to a smaller maximum expansion angle. The choice between airfoils is ultimately based on shock tube conditions, and if an attached shock can be achieved, the pre-corner portion may be excluded from consideration.

\subsection{\label{Maxabili}Maximum expansion degree}
\begin{figure}
	\centering
	\includegraphics[width=0.45\textwidth]{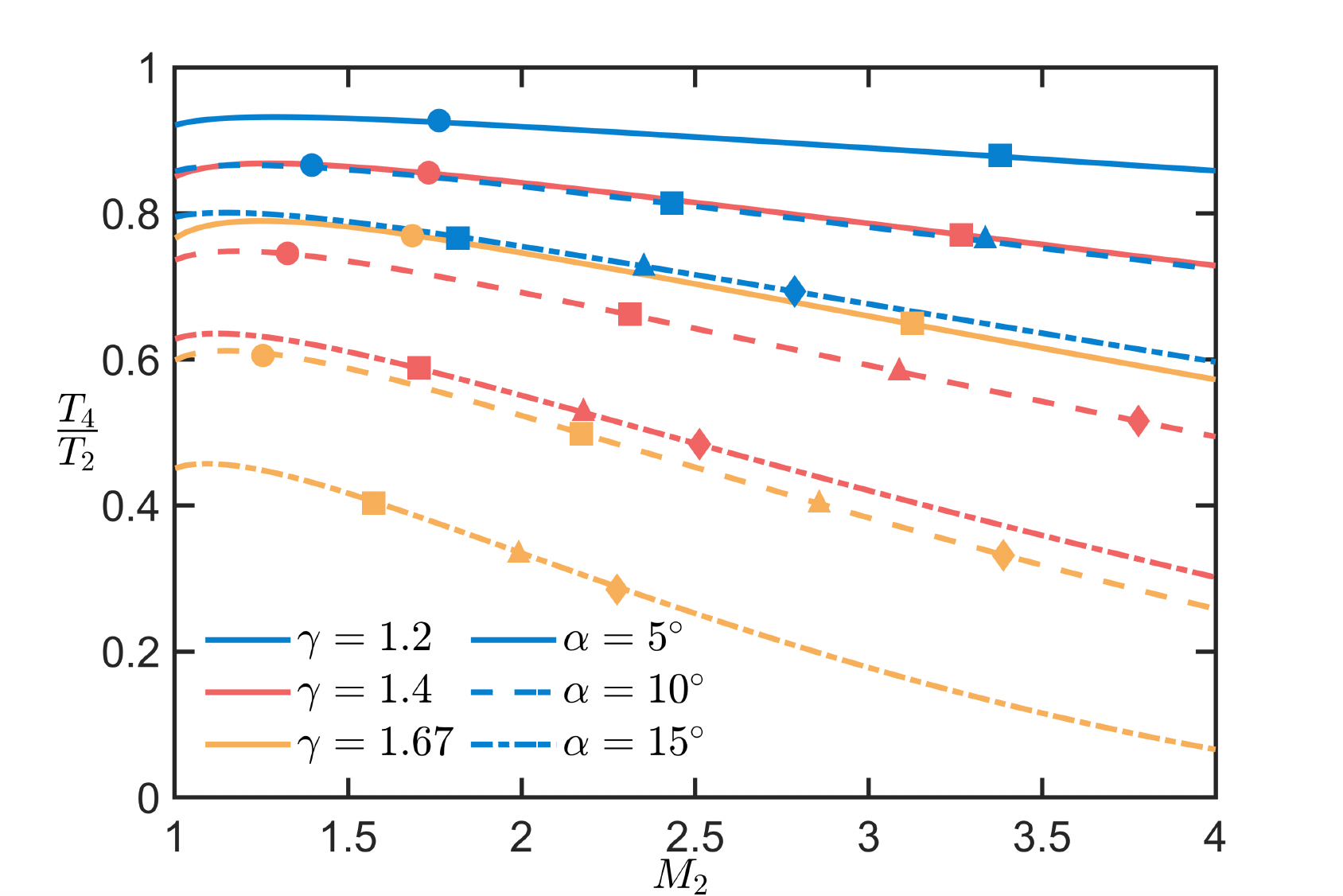}
	\caption{\label{fig:TempvsMa}Temperature ratio after crossing the double ideal Prandtl-Meyer expansion fans for different $\gamma$ and $\alpha$. The circle dot, square, triangle, and rhombus denote where $2L_\mathrm{tail}/D_\mathrm{af}=2,\;5,\;10,\; \mathrm{and}\; 20$, respectively.}
\end{figure}

The location of the tail of the Prandtl-Meyer expansion fans along the centerline ($L_\mathrm{tail}$) constrains the maximum expansion angle. A larger expansion region ($D_\mathrm{2}$) brings the flow closer to equilibrium after expansion and make PMD excessively large. The temperature ratio before and after expansion, for various specific heat ratios ($\gamma$) and expansion angles, is plotted in Fig.~\ref{fig:TempvsMa}. Both specific heat ratios and expansion angles significantly influence the temperature ratio, with the ratio decreasing as $\gamma$ and the expansion angle increase. Assuming the trajectories of expansion waves propagate along straight lines, $L_\mathrm{tail} = D_\mathrm{1} + D_\mathrm{2} = \frac{D_\mathrm{af}}{2}\cot\theta_\mathrm{2}$. The value of $L_\mathrm{tail}$ is determined by the combined effect of several parameters, with different $2L_\mathrm{tail}/D_\mathrm{af}$ values indicated on each line in Fig.~\ref{fig:TempvsMa}. 

$L_\mathrm{tail}$ increases rapidly with freestream Mach number and expansion angles, while specific heat ratios have a lesser impact. To limit $2L_\mathrm{tail}/D_\mathrm{af}$ smaller than 10, the expansion angle should be less than $15^\circ$ for freestream Mach numbers greater than 2.0, 2.2, and 2.4 for $\gamma$ = 1.2, 1.4, and 1.67, respectively. This constraint results in minimum temperature ratios around 0.7, 0.5, and 0.3, respectively. Consequently, achieving a significant decrease in temperature ratio becomes challenging when the specific heat ratio is low. When $2L_\mathrm{tail}/D_\mathrm{af}$ is very large, reducing $D_\mathrm{af}$ is an effective way to minimize $L_\mathrm{tail}$. However, smaller $D_\mathrm{af}$ reduces the height of the undisturbed zone, as shown in Fig.~\ref{fig:UndisZone}, affecting duct placement and radiance measurements.

\section{\label{DesignoP}Design of the Prandtl-Meyer plus duct arrangement}
\subsection{\label{Overview}Overview}
This section discusses the primary process of designing PMD for high-enthalpy radiating flow. The methodology involves theoretical analysis and CFD simulations to achieve a detailed design. Technically, PMD can be used in most shock tube conditions, provided equilibrium is achieved at the shock tube outlet and appropriate diagnostic techniques are employed. Different combinations of shock tube conditions and airfoil geometry result in distinct flow characteristics. Ideally, the gas should remain in a thermochemically frozen state when crossing the double Prandtl-Meyer expansions. However, complete freezing at the head of the expansion zone ($D_\mathrm{2}$) is challenging due to the high temperature behind the primary shock wave, which significantly enhances particle collision frequencies. Since chemical equilibrium typically takes longer than thermal equilibrium\cite{Rich1970}, the flow characteristics after crossing the expansion can be categorized as thermochemical equilibrium, thermal equilibrium with chemical freezing, thermochemical nonequilibrium, and thermochemical freezing.

To avoid the complexities associated with thermochemical nonequilibrium in the freestream, it is crucial to first assess the relaxation distance behind the incident shock wave. If equilibrium can be achieved, the next step involves the determination of the shock tube conditions and airfoil geometry. An analysis of the feasibility of the modified airfoil is first presented, followed by a strategy for designing the airfoil discussed in Section \ref{DeteroSTaA}. Subsequently, when the duct is employed, a solution to mitigate the leading-edge shock is provided. The final step involves determining the length of the effective measurement zone, which is influenced by shock tube conditions, flow establishment time, and transition onset location inside the duct.

\subsection{\label{Preset}Preset of the shock tube conditions}
To ensure the feasibility of the design conditions, it is important to have a comprehensive understanding of the shock tube capability. For a given shock tube conditions, the relaxation distance can be predicted by one-dimensional shock wave simulations, and the corresponding separation distance (\textit{l}) between the contact surface and the incident shock can be calculated using Mirels’ theory. The length of the equilibrium flow is the difference between the two parameters. Equilibrium can be typically achieved several centimeters downstream of the shock under high-enthalpy conditions, rendering \textit{l}\textsubscript{rel} negligible compared to \textit{l}. However, this may not hold for short shock tubes and low fill pressure. To address this problem, the relaxation distance (\textit{l}\textsubscript{rel}) can be decreased by increasing the shock velocity (\textit{V}\textsubscript{s}) or the fill pressure (\textit{P}\textsubscript{1}). Additionally, according to Mirels’ theory, the test gas length can be increased by reducing the shock velocity or increasing the fill pressure.
\begin{figure}
	\centering
	\includegraphics[width=0.4\textwidth]{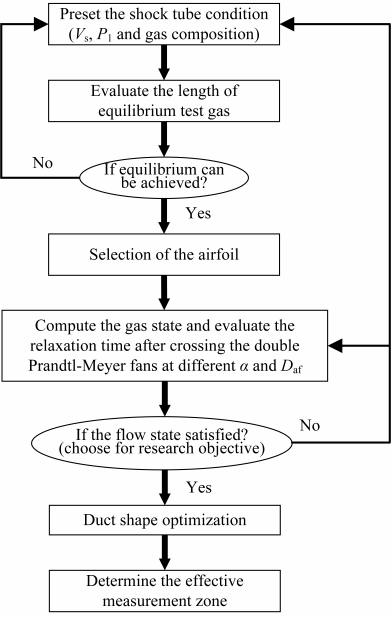}
	\caption{\label{fig:Flowchart}Flowchart of the design method of the Prandtl-Meyer plus duct arrangement.}
\end{figure}

The determination of the shock velocity is crucial and should align with the research objective; thus, large modifications to it are unnecessary. To generate high-enthalpy radiating flow, the shock velocity is typically high to introduce significant vibrational and electronic energy excitation. The fill pressure plays a crucial role in the relaxation time. The gas state behind the expansion and the length of the undisturbed zone are both significantly influenced by $\alpha$ and $D_\mathrm{af}$. Therefore, these factors must be carefully considered together for selecting appropriate shock tube conditions.

As mentioned earlier, PMD can theoretically achieve four types of flow states. Without considering the radiance intensity, the analysis in Section~\ref{QualiStudy} suggests that enlarging the undisturbed zone requires increasing $\alpha$ and $D_\mathrm{af}$ as much as possible while ensuring the desired flow conditions is maintained. It is worth noting that the same flow state can be achieved with both high and low fill pressures under the same shock velocity, with the former resulting in a larger $\alpha$ and thus a larger undisturbed zone. Therefore, for a satisfactory design, the shock tube should operate at a high fill pressure. In addition, according to Mirels’ theory\cite{Mirels1963}, increasing the fill pressure at a constant shock velocity leads to a longer test time, further emphasizing the advantage of high fill pressure conditions.

Designing a flow in thermal or chemical equilibrium is simpler compared to a nonequilibrium flow. This is because high fill pressure and large $D_\mathrm{af}$ facilitate the flow to reach equilibrium. The value of $\alpha$ should be as large as possible, as long as the flow remains in equilibrium. Achieving chemical equilibrium requires a high collision frequency, which is challenging due to the rapid temperature decrease within the small expansion distance ($D_\mathrm{2}$). Therefore, in most cases, the chemical reactions are nonequilibrium or even frozen. However, the vibrational and electronic energy can be in equilibrium as long as the fill pressure is sufficiently high. In this case, the recombination process is decoupled from the vibrational and electronic nonequilibrium. The radiance under equilibrium conditions is typically high due to the high particle number density resulting from the high fill pressure.

Thermochemical nonequilibrium or frozen conditions can be achieved by reducing the fill pressure and $D_\mathrm{af}$, or by increasing $\alpha$. When the fill pressure is high, a large $\alpha$ is necessary. This approach offers the advantage of obtaining a larger undisturbed zone. However, increasing $\alpha$ may lead to a detached shock, lower temperature, and reduced particle number density, which can pose challenges for radiance measurement. Therefore, it is recommended to use a small preset fill pressure that allows for the observation of thermochemical nonequilibrium after the expansion. The preset value can be initially based on experience. The main problem with this strategy is the low radiance intensity; therefore, subsequent modifications are required based on the evaluation of the preset conditions.

\subsection{\label{DeteroSTaA}Determination of the shock tube conditions and the airfoil}
\subsubsection{\label{SeleA}Selection of the airfoil}
\begin{figure}
	\centering
	\includegraphics[width=0.45\textwidth]{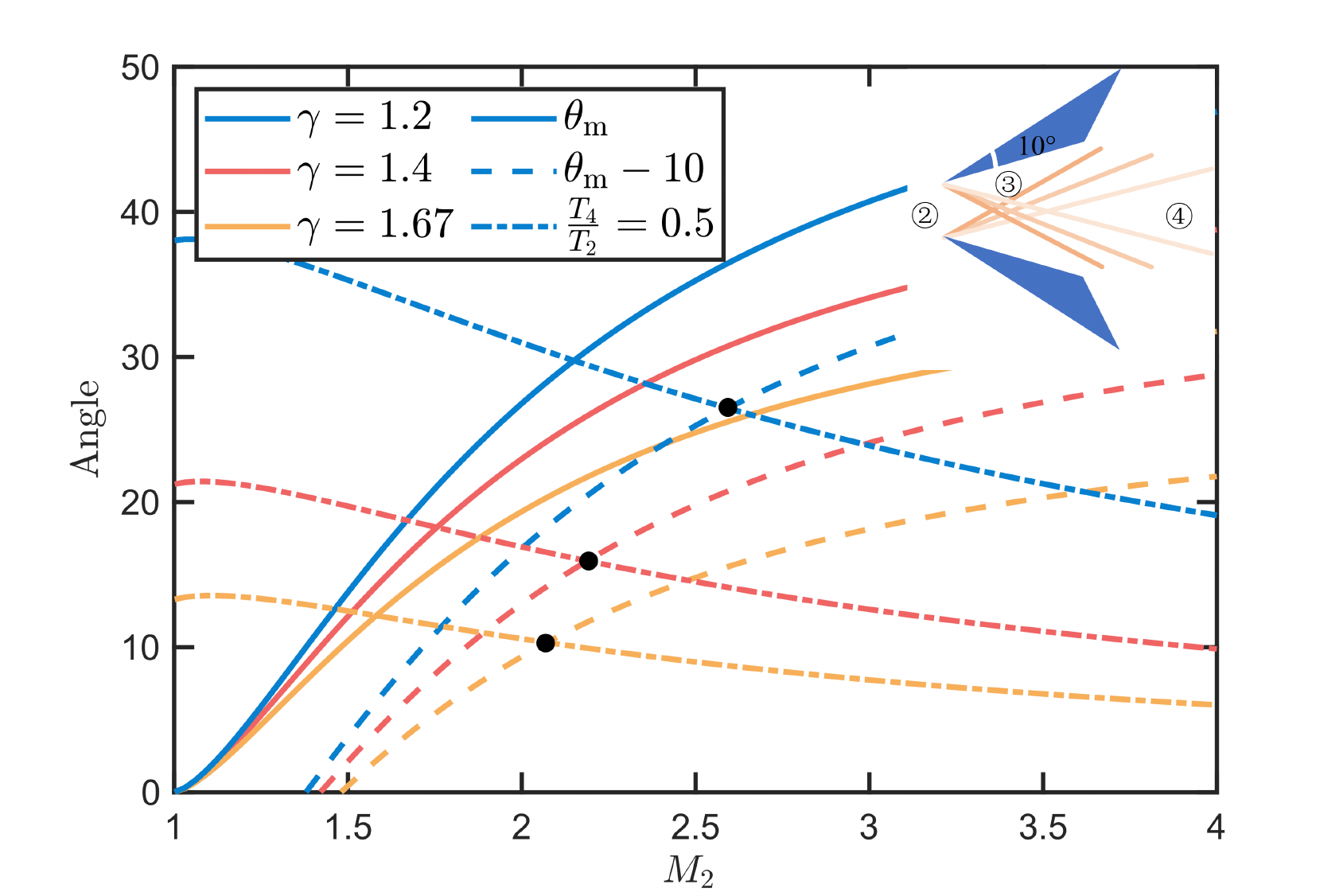}
	\caption{\label{fig:MExpAn}Comparison of maximum turn angles for an attached shock and expansion angles needed for a large decrease in temperature under different $\gamma$. Black circle dots indicate the critical Mach numbers to avoid a detached shock for different $\gamma$}
\end{figure}

Once the shock tube conditions are set, assessing the feasibility of using the modified airfoil is necessary. The maximum deflection angle depends on the specific heat ratio and Mach number. For ideal gas, the maximum deflection angle for an attached shock and the expansion angle required for a significant temperature decrease ($T_\mathrm{4}/T_\mathrm{2}=0.5$) are calculated and plotted in Fig.~\ref{fig:MExpAn} for different $\gamma$ and Mach numbers. The thickness of the modified airfoil must be sufficiently thin to avoid a detached shock, which also increases the risk of bending. It is found that an airfoil with $\beta$ equals to 10° is sufficiently thick to withstand the high driver pressure without deformation. The minimum Mach numbers are marked by the black circle dots for different $\gamma$ in Fig.~\ref{fig:MExpAn}. The Mach number required to achieve a significant temperature decrease increases as $\gamma$ decreases.

\subsubsection{\label{GasState}Gas state crossing the double Prandtl-Meyer expansion fans}
As discussed in Section~\ref{ModitoWil}. The determination of the modified airfoil geometry involves five parameters, with the expansion angle ($\alpha$) being the most crucial one, Therefore, it should be evaluated first. For gases with smaller $\alpha$, such as carbon dioxide, a larger $\alpha$ is required to achieve pronounced de-excitation. Conversely, for monatomic gases like helium, a smaller 
$\alpha$ is sufficient to cool the gas effectively. Typically, $\alpha$ can be set to a value that induces a significant temperature decrease, as discussed in the previous section. Further adjustments can be made based on evaluations of the gas state as it crosses the double Prandtl-Meyer expansion fans using CFD. 

In most cases, the flow is not entirely thermochemically frozen, and there is no straightforward method to accurately determine the gas state crossing the double Prandtl-Meyer expansions. Therefore, CFD is recommended to aid in the design process. The computational domain, depicted in Fig.~\ref{fig:GridsA}, adequately fulfills this requirement. To create a pronounced de-excitation, the expansion angle must be sufficiently large to lead to a noticeable temperature decrease and can be referred from Fig.~\ref{fig:MExpAn}, using the Mach number and frozen $\gamma$ behind the shock. Then, the preset shock tube conditions can be evaluated using the computational domain depicted in Fig.~\ref{fig:GridsA}. In addition to an appropriate thermochemical model, an inviscid model can also be applied due to the minimal influence of the viscous effect on the centerline flow. 

\begin{figure}
	\centering
	\includegraphics[width=0.45\textwidth]{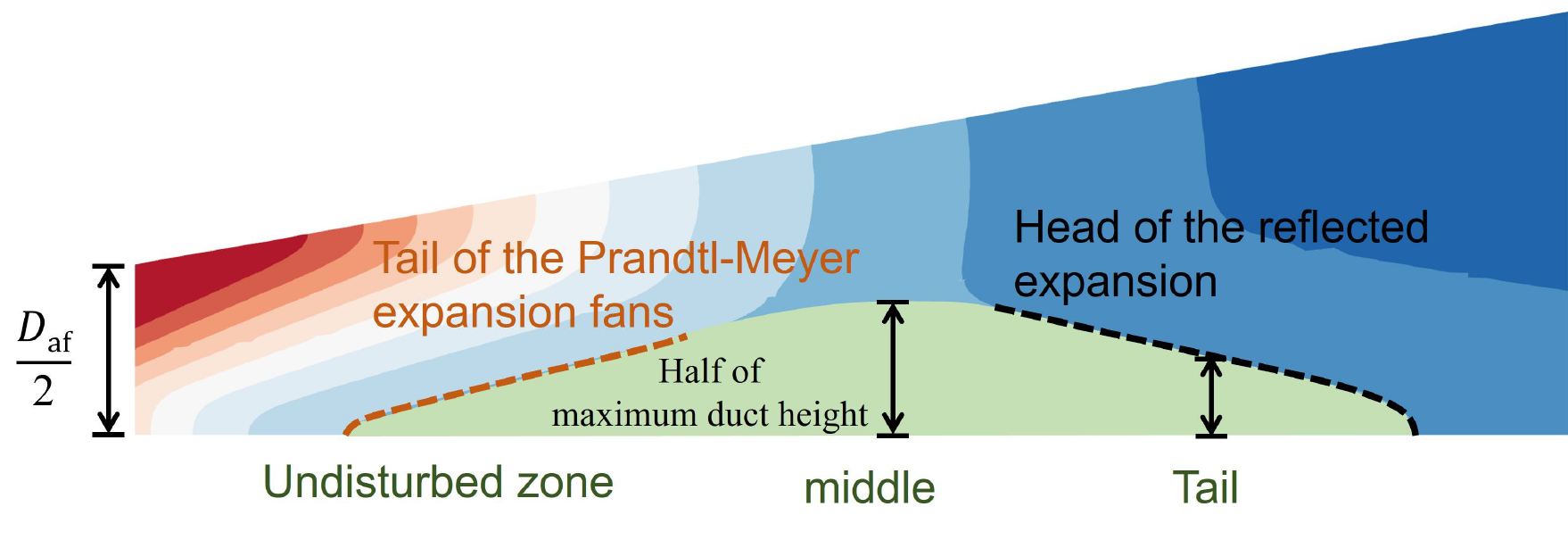}
	\caption{\label{fig:UndisZone}Two-dimensional structure of the undisturbed zone.}
\end{figure}

In this domain, the flow state behind the expansion is also influenced by $D_\mathrm{af}$, as discussed by Slack\cite{Slack1969}, because the expansion distance on the centerline ($D_\mathrm{2}$) is also determined by $D_\mathrm{af}$. A smaller $D_\mathrm{af}$ will result in a flow state closer to nonequilibrium and narrow the undisturbed zone in the vertical dimension, affecting the allowable height of the duct. The height of the undisturbed zone is determined by $\alpha$, $D_\mathrm{af}$, and duct position. Fig.~\ref{fig:UndisZone} illustrates the two-dimensional structure of the undisturbed zone, which lies between the tail of the Prandtl-Meyer expansion fans and the head of the reflected expansion. The duct can be positioned anywhere within this zone, with the maximum height found in the middle. Since $\alpha$ is determined, the adjustment of the size of the undisturbed zone should be achieved by changing $D_\mathrm{af}$. To measure the radiating flow on the centerline, the duct height should be sufficiently large, with a critical value of 10 mm set in this study. For high-enthalpy conditions, it is recommended to place the duct at the tail of the undisturbed zone, since the time required for flow establishment inside the duct is comparable to the short shock tube test time, resulting in a short measurement zone inside the duct. Additionally, the diameter of the shock tube also imposes limitations on $D_\mathrm{af}$. In equilibrium conditions, the preset $D_\mathrm{af}$ can exceed that of nonequilibrium conditions due to its influence on the expansion distance. Typically, it can be set to the maximum value allowed by the shock tube diameter. For nonequilibrium conditions, a small preset value is required, which can be around twice the critical value of the duct height (20 mm). Further adjustment of $D_\mathrm{af}$ should be considered when the simulated maximum height of the undisturbed zone is small.

As discussed in Section~\ref{QualiStudy}, the preset $L_\mathrm{af}$ must be larger than $x_\mathrm{p}$ to maximize the undisturbed zone. Accurate calculation of the reflection location requires the method of characteristics to compute the interference region, which consists of type $C_\mathrm{+}$ and type $C_\mathrm{-}$ characteristic lines, as shown in Fig.~\ref{fig:CharaCompAft}. The blue dashed characteristic line, originating from the leading edge of the lower airfoil, intersects the interference region and changes its direction of propagation, eventually becoming the dashed green line and reflecting on the lower wall of the upper airfoil. The gas properties behind the expansion can be determined under frozen assumption, thereby allowing the gradients of these two lines to be calculated. Using the geometric relationships, an approximation of the reflection point location can be obtained by computing the average gradient of the two lines (blue and green dashed lines):
\begin{equation}
	x_\mathrm{p} = \frac{D_\mathrm{af}}{(\tan\mu_1 + \tan\theta_2)/2 - \tan\alpha},
	\\
	\quad y_\mathrm{p} = x_\mathrm{p} \tan\alpha
\end{equation}
It can be observed that the reflection location depends on $D_\mathrm{af}$ and $\alpha$, where an increase in $D_\mathrm{af}$ and $\alpha$ results in an increase in $x_\mathrm{p}$ and $y_\mathrm{p}$. Therefore, the determination of Laf should follow the determination of $D_\mathrm{af}$ and $\alpha$. Since the accurate location of the reflection point can only be obtained from CFD simulations, the preset Laf can be 1.5 times of $x_\mathrm{p}$. Generally, A larger $L_\mathrm{af}$ is recommended for a conservative design to maintain the undisturbed zone interrupted by the reflected expansion instead of the corner expansion.

Once the gas properties behind the expansion are obtained, radiance from the high-enthalpy flow can be calculated using the absorption and emission coefficients, enabling a preliminary evaluation of the feasibility of OES measurement. Based on these evaluations for different shock tube conditions, further modifications can be made to the shock tube conditions and airfoil geometry. For thermodynamic equilibrium conditions, the typical modification is to find the maximum $\alpha$ to maintain equilibrium, since preset fill pressure and $D_\mathrm{af}$ approach their maximum values. For nonequilibrium conditions, fill pressure, $\alpha$ and $D_\mathrm{af}$ should be considered altogether. Increasing $\alpha$, decreasing $D_\mathrm{af}$, and lowering the fill pressure bring the flow closer to nonequilibrium, while these changes also result in shorter shock tube test times and lower particle number densities, which affects the radiance intensity. Table~\ref{tab:SummaryotI} lists a summary of the influence of shock tube conditions and airfoil geometry. Researchers must have a comprehensive understanding of which aspect of modification is more important under the current design.

The location of the curved shock, as shown in Fig.~\ref{fig:ModiDeg}, is determined by $L_\mathrm{af}$, $L_\mathrm{af2}$, and $\beta$. They must be sufficiently large to prevent the curved shock from reaching the centerline before the reflected expansion while being small enough for $\beta$ to avoid a detached shock at the leading edge of the airfoil. Additionally, the airfoil must be thick enough to withstand the pressure difference on the upper and lower surfaces without deformation. Accurate determination of $L_\mathrm{af2}$ also requires CFD simulations, although it can be simply set to a large value. Typically, $L_\mathrm{af2}=L_\mathrm{af}/2$ is sufficiently large to avoid the convergence of the corner expansion and curved shock. Generally, the maximum undisturbed zone varies for each combination of shock tube conditions and airfoil geometry and is limited by the reflection of Prandtl-Meyer expansion waves. $L_\mathrm{af2}$ and $\beta$ are used to prevent the convergence of the corner shock wave and the reflected expansion wave. 
\begin{table*}
 \renewcommand{\arraystretch}{1.2}
	\caption{\label{tab:SummaryotI}Summary of the influence of shock tube conditions and airfoil geometry.}
	\begin{ruledtabular}
		\begin{tabular}{lcccc}
			Parameters&Relaxation&Undisturbed zone&Test time&Number density\\
			\hline
			\textit{V}\textsubscript{s} &Towards equilibrium &/ 
			&Decrease& More dissociation and Ionization
			\\
			\textit{P}\textsubscript{1}  &Towards equilibrium &/ 
			&Increase      & Increase	
			\\
			$\alpha$  &Towards frozen & Enlarge 
			&/      & Decrease	
			\\
			$D_\mathrm{af}$  &Towards equilibrium & Enlarge 
			&/      & /
			\\
		\end{tabular}
	\end{ruledtabular}
\end{table*}

\subsection{\label{DuctSO}Duct shape optimization}
\begin{figure}
	\centering
	\includegraphics[width=0.45\textwidth]{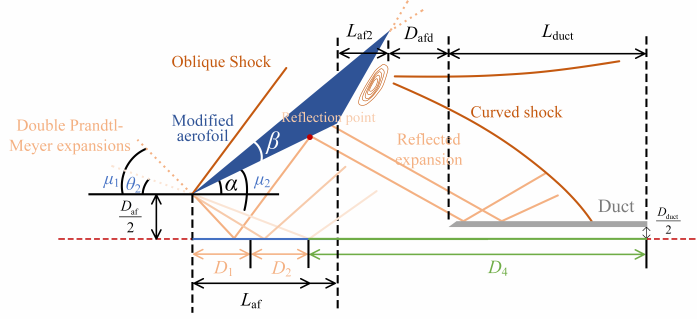}
	\caption{\label{fig:PrimaryWD}Primary wave diagram with modified airfoil when a constant-area duct is added.}
\end{figure}

A constant-area duct, shown in Fig.~\ref{fig:PrimaryWD}, can be placed at the undisturbed zone to prevent external disturbances from disrupting the steady centerline flow. In cases of high Mach number flow over a flat plate, the strong viscous interaction leads to the generation of a shock at the leading edge, which continuously reflects on both side of the duct, disrupting the the uniformity of the centerline flow. Therefore, modifications to the duct shape are required and can be based on the calculation of the displacement thickness, which is determined by solving the compressible boundary layer equations under the frozen or equilibrium assumption\cite{Anderson2006}. The displacement thickness can be computed by
\begin{equation}
	\delta^{*} = \int_{0}^{\delta} \left(1 - \frac{\rho u}{\rho_\mathrm{e}u_\mathrm{e}}\right) dy
\end{equation}
where $\rho_\mathrm{e}$ and $u_\mathrm{e}$ are freestream density and velocity, respectively. However, the boundary layer is typically in thermochemical nonequilibrium, resulting in a displacement thickness that lies between the values for a frozen or equilibrium flow. One can compute the frozen or equilibrium value and then scale it accordingly. Since the inflow into the duct is already known from previous simulations, this simulation can be carried out solely with the duct until an appropriate curve is obtained.

\subsection{\label{DetEMZ}Determination of the effective measurement zone}
\begin{figure}
	\centering
	\includegraphics[width=0.45\textwidth]{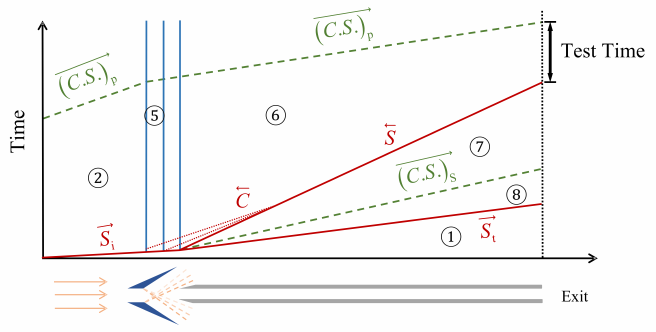}
	\caption{\label{fig:Theoreticalxt}Theoretical $x-t$ diagram for the shock tube flow in the duct. $\overrightarrow{S_\mathrm{t}}$ : transmitted shock; $\overrightarrow{(C.S.)_\mathrm{s}}$: secondary contact surface; $\overleftarrow{C}$, $\overleftarrow{S}$: rearward facing compression shock; $\overrightarrow{(C.S.)_\mathrm{p}}$: primary contact surface; $\overrightarrow{S_\mathrm{i}}$: incident shock. (adapted from Ref. \onlinecite{Slack1969})}
\end{figure}

The effective measurement zone of PMD is determined by three aspects: the effective test time, the transition onset location, and the radiance intensity. The flow establishment process of PMD was initially analyzed by Slack \cite{Slack1969}, and the theoretical $x\mbox{-}t$ diagram is shown in Fig.~\ref{fig:Theoreticalxt} As the transmitted shock, with a lower Mach number compared to the incident shock wave, passes through the duct, a secondary contact surface is formed to separate the gas downstream of the expansion from the upstream flow. Since the pressure behind the incident shock is greater than that which has crossed the Prandtl-Meyer expansion waves, several compressions are generated between zones (6) and (7), eventually coalescing into a shock. Following the rearward shock, the flow remains stable until the primary contact surface is reached. Therefore, the effective test time for different locations within the duct is the interval between the rearward shock and the primary contact surface, denoted as zone 6 in Fig.~\ref{fig:Theoreticalxt}.

The flow establishment time can be estimated by
\begin{equation}
	t_\mathrm{e} \approx K \frac{L_\mathrm{e}}{U_{\infty}}
\end{equation}
where $L_\mathrm{e}$ is the characteristic length, $U_\mathrm{\infty}$ is the freestream velocity, and $K$ is a constant determined by the flow phenomenon. Therefore, when the duct is added, although the undisturbed zone can be significantly extended, the effective test may be limited for locations further from the duct front.

\section{\label{AExample}An Example of High-enthalpy Radiating N\textsubscript{2} Flows}
This section presents a design case for reference, with pure N\textsubscript{2} as the test gas and OES being the diagnostic method. The research objective of this case is to study the nonequilibrium de-excitation. To achieve a significant amount of electronic excitation, a shock velocity of 7 km/s is set, which induces approximately half the dissociation of N\textsubscript{2}. The radiance of interest in this example is the molecular band radiation in the UV-visible spectral region (300-450 nm), where the main transitions are N\textsubscript{2}\textsuperscript{+} first negative and N\textsubscript{2} second positive systems.
\subsection{\label{ExmPoST}Preset of the shock tube conditions}
The JF-10 shock tunnel\cite{Zhao2005}, which can also be modified into a shock tube, is selected for designing the shock tube conditions. The driver and driven section lengths are 10.2 m and 12.5 m, respectively, with identical inner diameters of 150 mm. To account for shock attenuation caused by the viscous effect, a higher shock velocity (15\% higher than 7 km/s) is used as input for Mirels’ theory. Two different fill pressures, 20 Pa and 500 Pa, are chosen for the preliminary evaluation. The fill pressure of 20 Pa is set for studying vibrational and electronic de-excitation. The fill pressure of 500 Pa is close to the maximum capability of JF-10 for a shock velocity of 8 km/s (15\% greater than 7 km/s under inviscid simulation) and is designed for recombination measurement under thermal equilibrium. The shock tube test time and the test gas length predicted by Mirels’ theory are listed in Table~\ref{tab:shockWC} as conditions 1 and 2 with primary shock wave conditions. 

\begin{figure*}
	\subfigure[Condition 1, temperature]{\label{fig:distributionoNA}\includegraphics[width=0.45\textwidth]{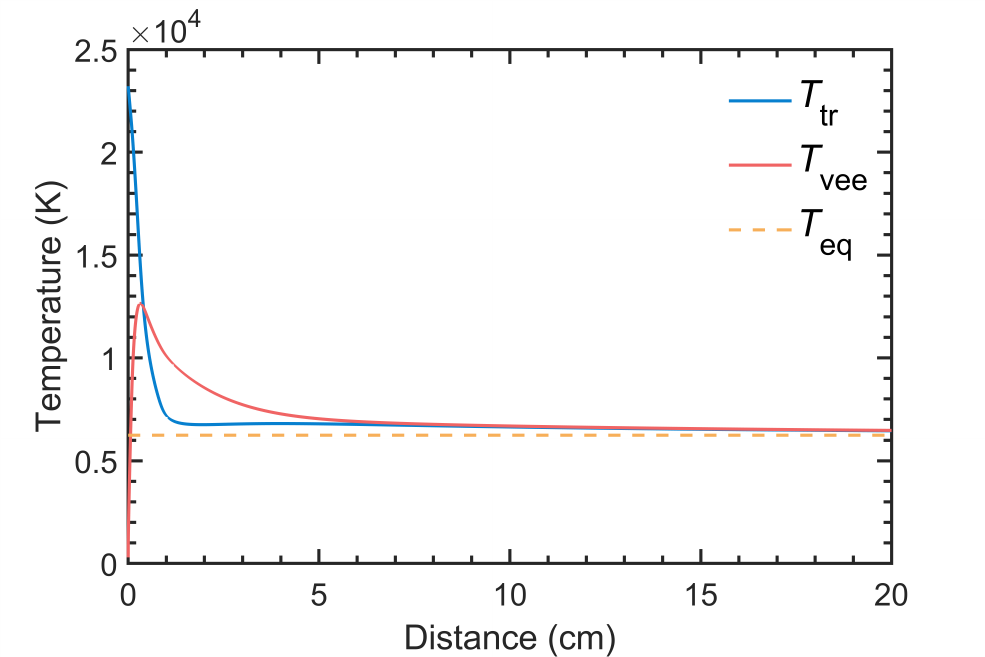}}
	\subfigure[Condition 1, mass franctions]{\label{fig:distributionoNB}\includegraphics[width=0.45\textwidth]{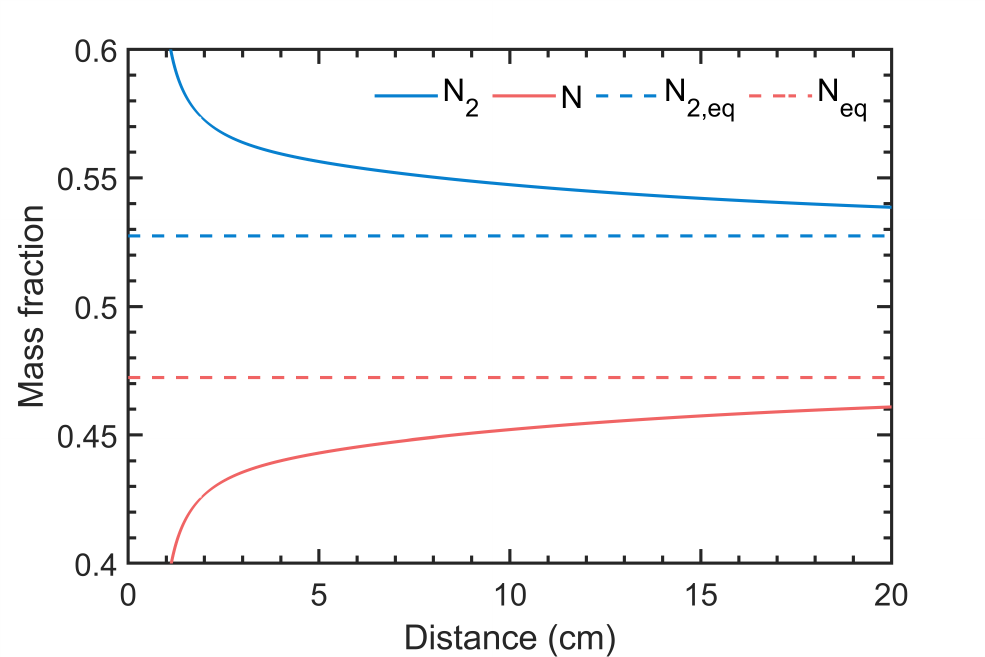}}
	\subfigure[Condition 3, temperature]{\label{fig:distributionoNC}\includegraphics[width=0.45\textwidth]{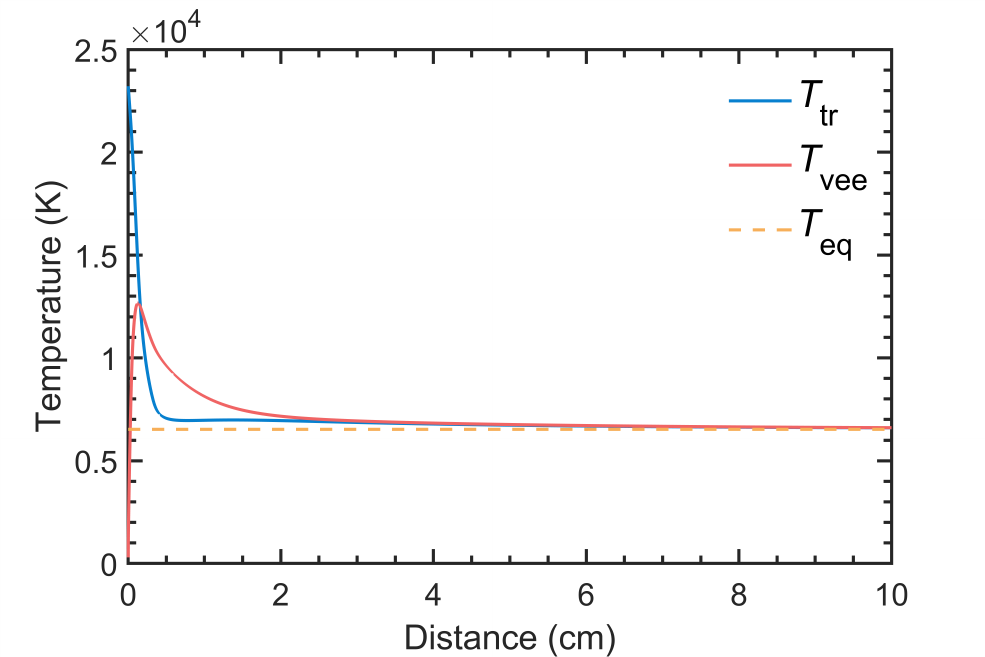}}
	\subfigure[Condition 3, mass fractions]{\label{fig:distributionoND}\includegraphics[width=0.45\textwidth]{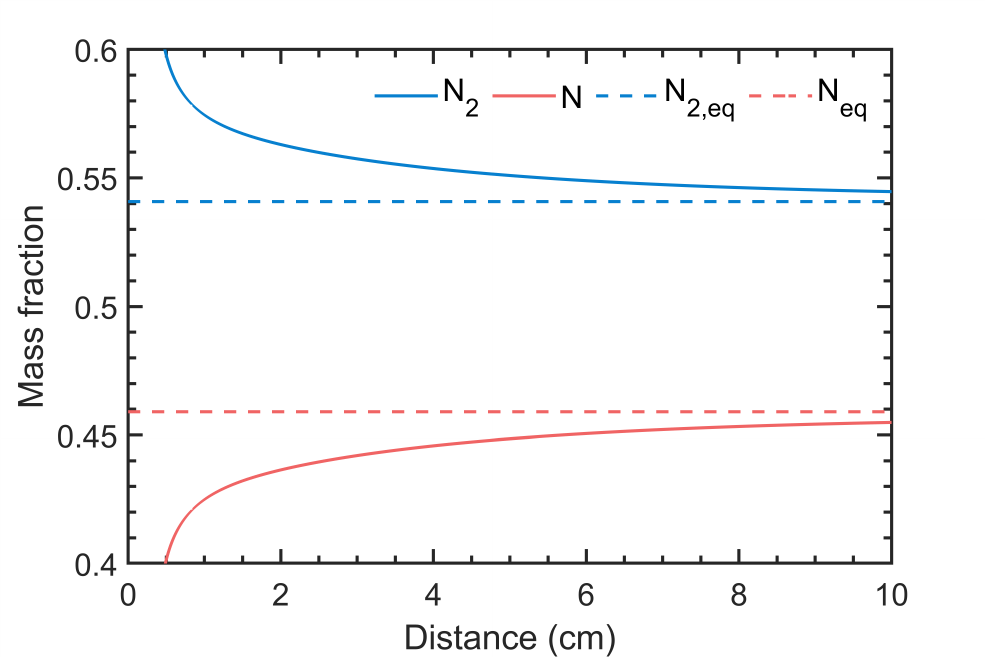}}
	\caption{\label{fig:distributionoN}Distributions of nonequilibrium temperatures and species mass fractions after the incident shock front under conditions 1 and 3. “eq” denotes the flow properties under thermochemical equilibrium.}
\end{figure*}

Condition 1, characterized by a lower fill pressure, requires a longer distance for the flow to reach thermochemical equilibrium. Therefore, it is examined as an example, and the distributions of nonequilibrium temperatures and species mass fractions are shown in Figs.\ref{fig:distributionoN}(a, b). The equilibrium gas properties are obtained from the SDToolbox\cite{Browne2018}. The flow is considered in equilibrium when only a 1\% difference in temperature and N\textsubscript{2}/N mass fractions is observed between the equilibrium calculation and the one-dimensional simulation. At around 10 cm, the flow reaches thermal equilibrium, while chemical nonequilibrium persists at $x=0.20\;\mathrm{m}$, which equals the test gas length predicted by Mirels’ theory (0.2 m). Therefore, increasing the fill pressure is necessary. The shock wave conditions of increasing fill pressure to 50 Pa is listed in Table~\ref{tab:shockWC} as condition 3, and the corresponding distributions of nonequilibrium temperatures and species mass fractions are displayed in Figs.\ref{fig:distributionoN}(c, d), with thermodynamic and chemical equilibrium achieved at \textit{x} = 0.04 m and \textit{x} = 0.08 m, respectively. Additionally, the test gas length increases to 0.33 m, ensuring sufficient equilibrium inflow for the experiment.

\begin{table*}
	 \renewcommand{\arraystretch}{1.5}
	
	\caption{\label{tab:shockWC}Shock wave conditions.}
	\begin{ruledtabular}
		\begin{tabular}{lccccccccc}
		  \multirow{2}{*}{\makecell{Conditions}}           
		& \multirow{2}{*}{\makecell{Shock\\ velocity\\\textit{V}\textsubscript{s} (m/s)}} 
		& \multirow{2}{*}{\makecell{Fill\\ pressure\\\textit{P}\textsubscript{1} (Pa)}} 
		& \multirow{2}{*}{\makecell{Initial gas\\ composition}} 
		& \multirow{2}{*}{\makecell{Enthalpy\\ (MJ/kg)}}   
		& \multirow{2}{*}{\makecell{Mach number \\behind the shock}} 
		& \multicolumn{2}{c}{\makecell{Mass fractions\\ behind the shock}} 
		& \multirow{2}{*}{\makecell{Test time\\$(\mathrm{\mu s})$}} 
		& \multirow{2}{*}{\makecell{Test gas\\ length (m)}} 
		\\\cline { 7 - 8 }
		&  &  &  &  &  & N\textsubscript{2}& N &  &  
		\\\hline
		1 & 7.0 & 20 & $\mathrm{N}_{2}$  & 24.41 & 3.20 & 0.53 & 0.47 & 24 & 0.20 
		\\
		2 & 7.0 & 500 & $\mathrm{N}_{2}$ & 24.39 & 3.40 & 0.58 & 0.42 & 80 & 0.61 
		\\
		3 & 7.0 & 50 & $\mathrm{N}_{2}$  & 24.39 & 3.34 & 0.54 & 0.46 & 41 & 0.33 
		\\
		4 & 7.0 & 100 & $\mathrm{N}_{2}$ & 24.40 & 3.30 & 0.55 & 0.45 & 54 & 0.42 
		\\
		\end{tabular}
	\end{ruledtabular}
\end{table*}

\subsection{\label{ExmDetoSTaA}Determination of the shock tube conditions and the airfoil}
\subsubsection{\label{ExmSeleA}Selection of the airfoil}
The Mach numbers behind the shock for both conditions (2 and 3) are around 3.30, surpassing the minimum value depicted in Fig.~\ref{fig:MExpAn}. Consequently, the modified airfoil is applicable under present conditions.

\subsubsection{\label{ExmGasState}Gas state crossing the double Prandtl-Meyer expansion fans}
\begin{figure*}
	\subfigure[Condition 3, \textit{p}\textsubscript{1}=50 Pa]
	{\label{fig:distributionoNtA}\includegraphics[width=0.45\textwidth]{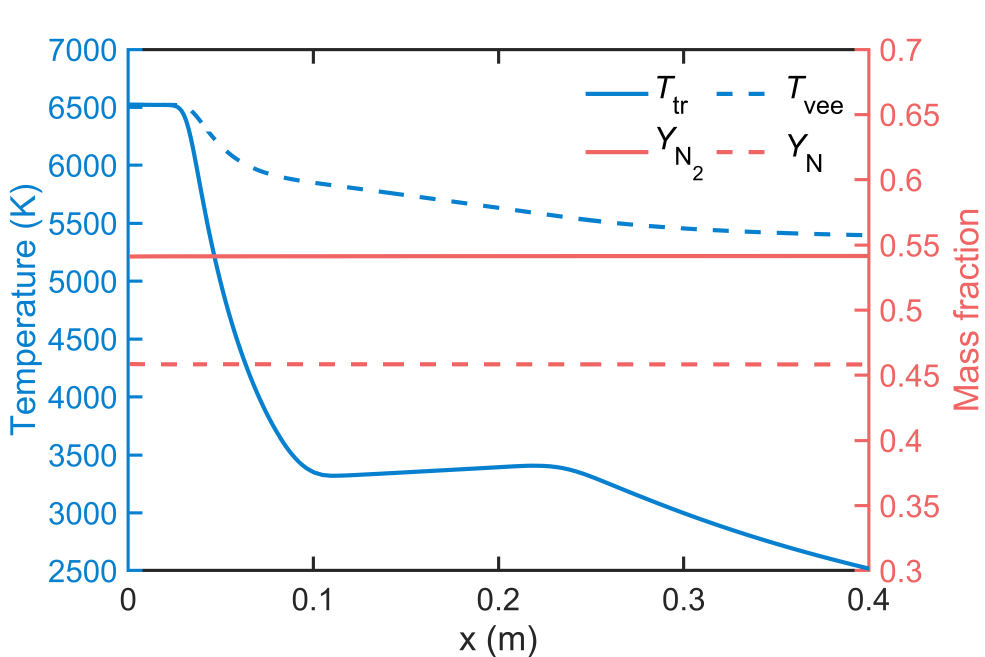}}
	\subfigure[Condition 2, \textit{p}\textsubscript{1}=500 Pa]
	{\label{fig:distributionoNtB}\includegraphics[width=0.45\textwidth]{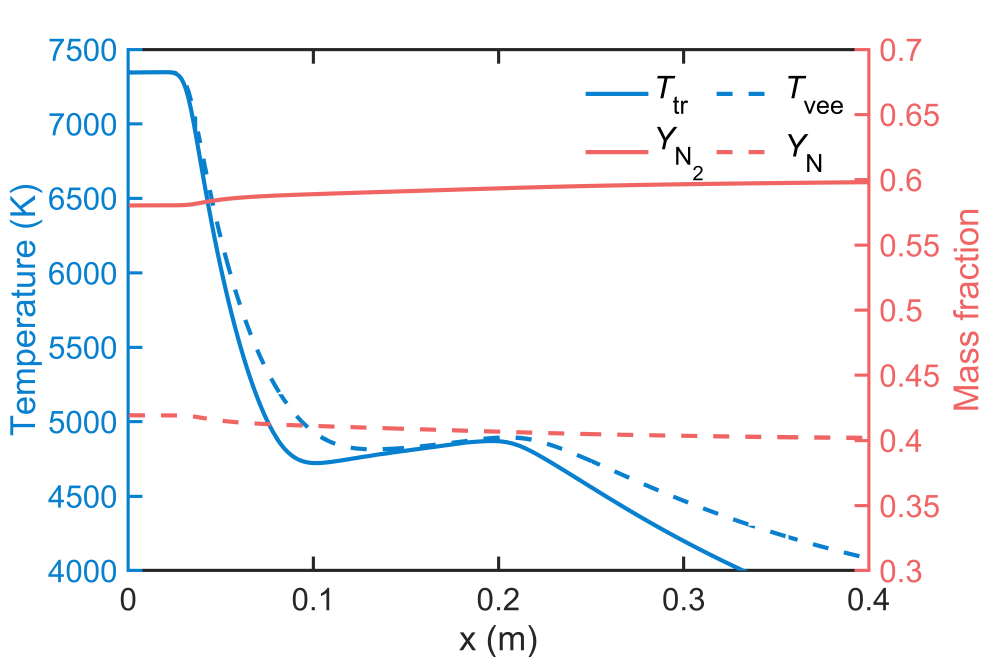}}
	\subfigure[Condition 4, \textit{p}\textsubscript{1}=100 Pa]
	{\label{fig:distributionoNtC}\includegraphics[width=0.45\textwidth]{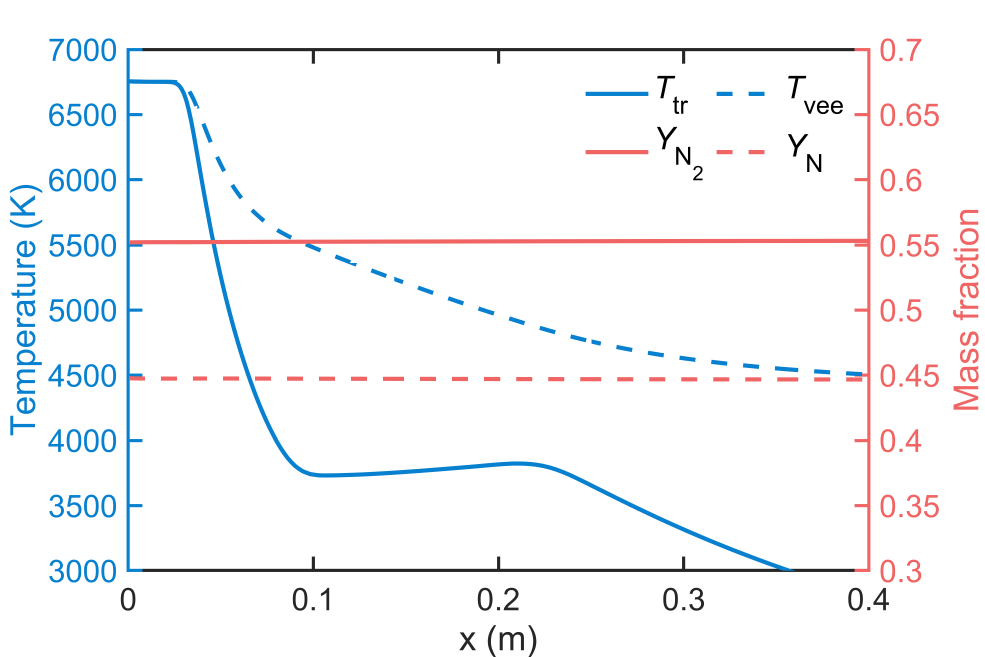}}
	\subfigure[Spectral radiance]{\label{fig:distributionoNtD}\includegraphics[width=0.45\textwidth]{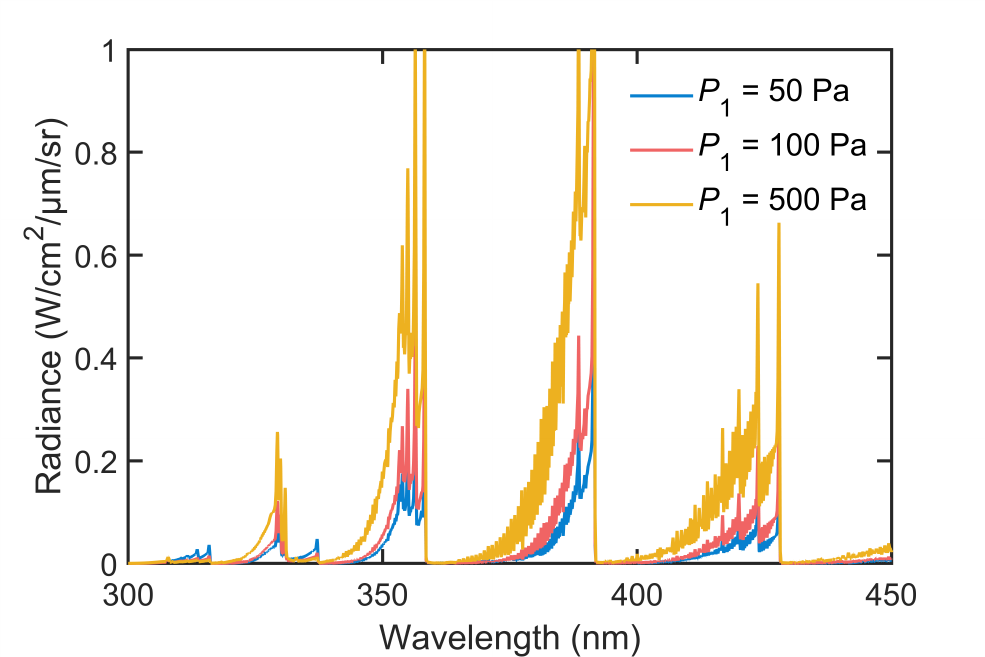}}
	\caption{\label{fig:distributionoNt}Distributions of nonequilibrium temperatures, mass fraction, and spectral radiance under different shock tube conditions.}
\end{figure*}

According to Fig.~\ref{fig:MExpAn}, an expansion angle of 10° is employed to induce a significant temperature drop. For conditions 2 and 3, $D_\mathrm{af}$ is set to 44 mm and 22 mm, respectively, which correspond to approximately 15\% and 30\% of the shock tube diameter. An identical $L_\mathrm{af}$ of 0.5 m is used for both conditions, which is three times larger than xp. Using the computational domain depicted in Fig.~\ref{fig:GridsA}, the simulation results are shown in Fig.~\ref{fig:distributionoNt}(a,b). \textit{x} = 0 m indicates the leading edge of the airfoil. As the fill pressure increases from 50 to 500 Pa, the flow behind the expansion changes from thermal nonequilibrium with chemical frozen to thermal equilibrium with chemical nonequilibrium. Consequently, condition 3 is suitable for studying vibrational and electronic de-excitation under chemical freezing, while condition 2 is suitable for studying atomic recombination under thermal equilibrium. The simulated spectral radiance is shown in Fig.~\ref{fig:distributionoNtD}. It can be observed that both conditions have strong radiance to facilitate the measurement.

For condition 3, further modification can be made by increasing the fill pressure, since a high fill pressure increases the particle density behind the expansion and the shock tube test time. Fig.~\ref{fig:distributionoNtC} illustrates the flow properties after increasing the fill pressure to 100 Pa, and this condition is included in Table~\ref{tab:shockWC} as condition 4. The flow remains in thermal nonequilibrium with chemical freezing along the centerline, with stronger radiance and longer shock tube test time. Condition 4 has a shorter test time and undisturbed zone than condition 2, making the measurement more challenging and requiring a more detailed evaluation, and the following design steps for both conditions are similar. Therefore, the subsequent design process of condition 4 in this example will be shown.

\begin{figure}
	\centering
	\includegraphics[width=0.45\textwidth]{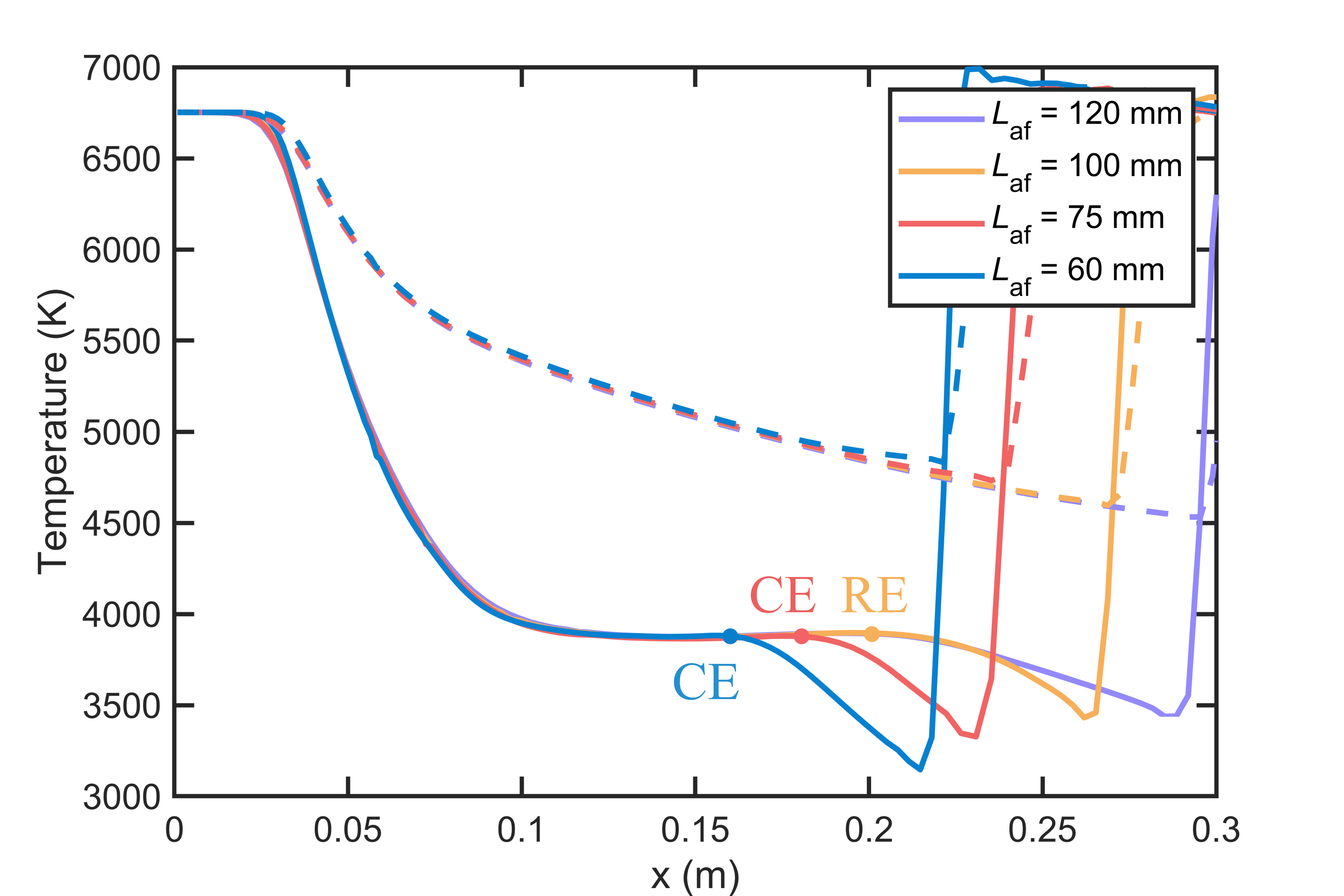}
	\caption{\label{fig:TemperatureDa}Temperature distribution along centerline when $\alpha$ = 10°, $D_\mathrm{af}$ = 22 mm. CE or RE indicates the undisturbed zone is interrupted by the corner or reflected expansion waves.}
\end{figure}

For condition 4, $\alpha$ and $D_\mathrm{af}$ are set to 10° and 22 mm, respectively, and the location of the reflection point is approximately at \textit{x} = 75 mm. The flow field around the airfoil with varying $L_\mathrm{af}$ was simulated to investigate the effect of increasing $L_\mathrm{af}$ on the undisturbed zone more distinctly, and the pressure distribution along the centerline is depicted in Fig.~\ref{fig:TemperatureDa}. The undisturbed zone enlarges when $L_\mathrm{af}$ increases from 60 to 100 mm and is interrupted by corner expansion waves. When $L_\mathrm{af}$ exceeds 100 mm, the size of the undisturbed zone remains unchanged and is interrupted by the reflected expansion waves. In summary, $\alpha$ = 10°, $D_\mathrm{af}$ =22 mm, and $L_\mathrm{af} > 100$ mm are suitable for investigating the electronic de-excitation. $L_\mathrm{af2}$ can be set to larger than 50 mm to avoid the convergence of the corner shock and corner expansion. It is important to note that the appropriate geometry of the airfoil depends on the shock tube conditions and the targeted flow characteristics.

\subsection{\label{ExmDuctSO}Duct shape optimization}
\begin{figure}
	\centering
	\includegraphics[width=0.45\textwidth]{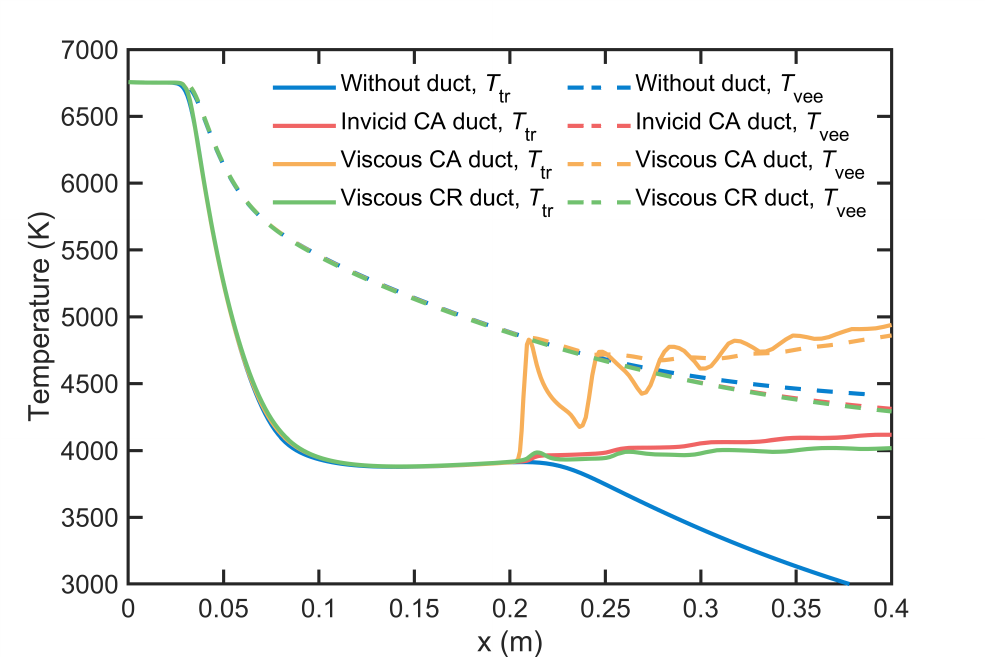}
	\caption{\label{fig:Temperaturepa}Comparison of temperature profiles along the centerline. CA and CR denote constant area and curved, respectively.}
\end{figure}

As shown in Fig.~\ref{fig:PrimaryWD}, a constant-area duct can be positioned in the undisturbed zone to capture the centerline flow, thereby extending its length. For this condition, the duct is recommended to be placed at the tail of the undisturbed zone due to the short shock tube test time. Inferring from the pressure distribution, a duct with a height of 10 mm would comfortably fit within the vertical space at \textit{x} = 0.19 m in the undisturbed zone. Fig.~\ref{fig:Temperaturepa} highlights the difference between simulations with inviscid and viscous models. In the viscous case, a pronounced viscous interaction is observed at the duct front, where the leading-edge shock repeatedly reflects off the inner wall, causing significant temperature fluctuations and leading to thermal equilibrium earlier compared to the inviscid scenario. Therefore, it is necessary to reconsider the duct shape to minimize the detrimental effects of the leading-edge shock.

\begin{figure}
	\subfigure[]{\label{fig:machndA}\includegraphics[width=0.45\textwidth]{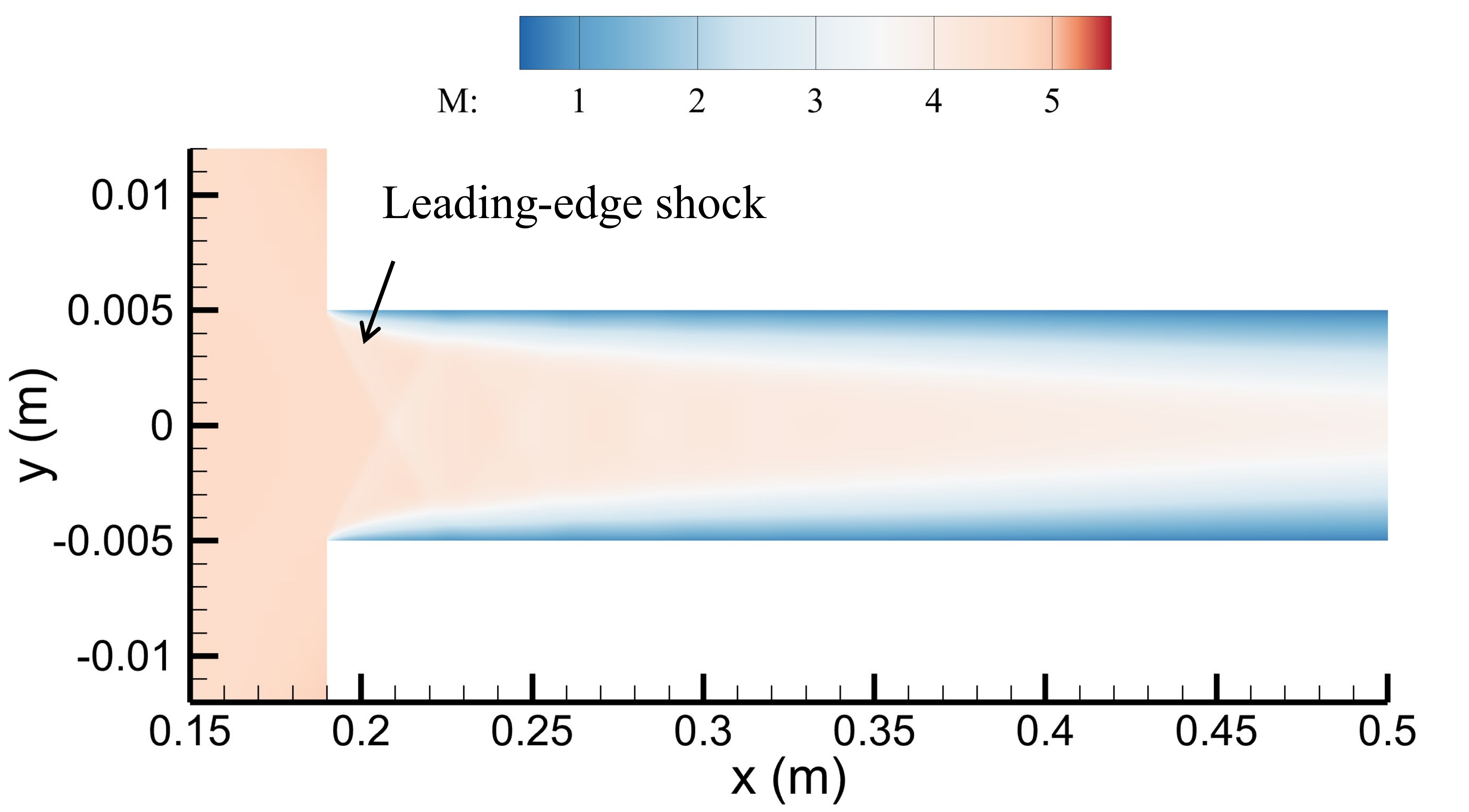}}
	\subfigure[]{\label{fig:machndtB}\includegraphics[width=0.45\textwidth]{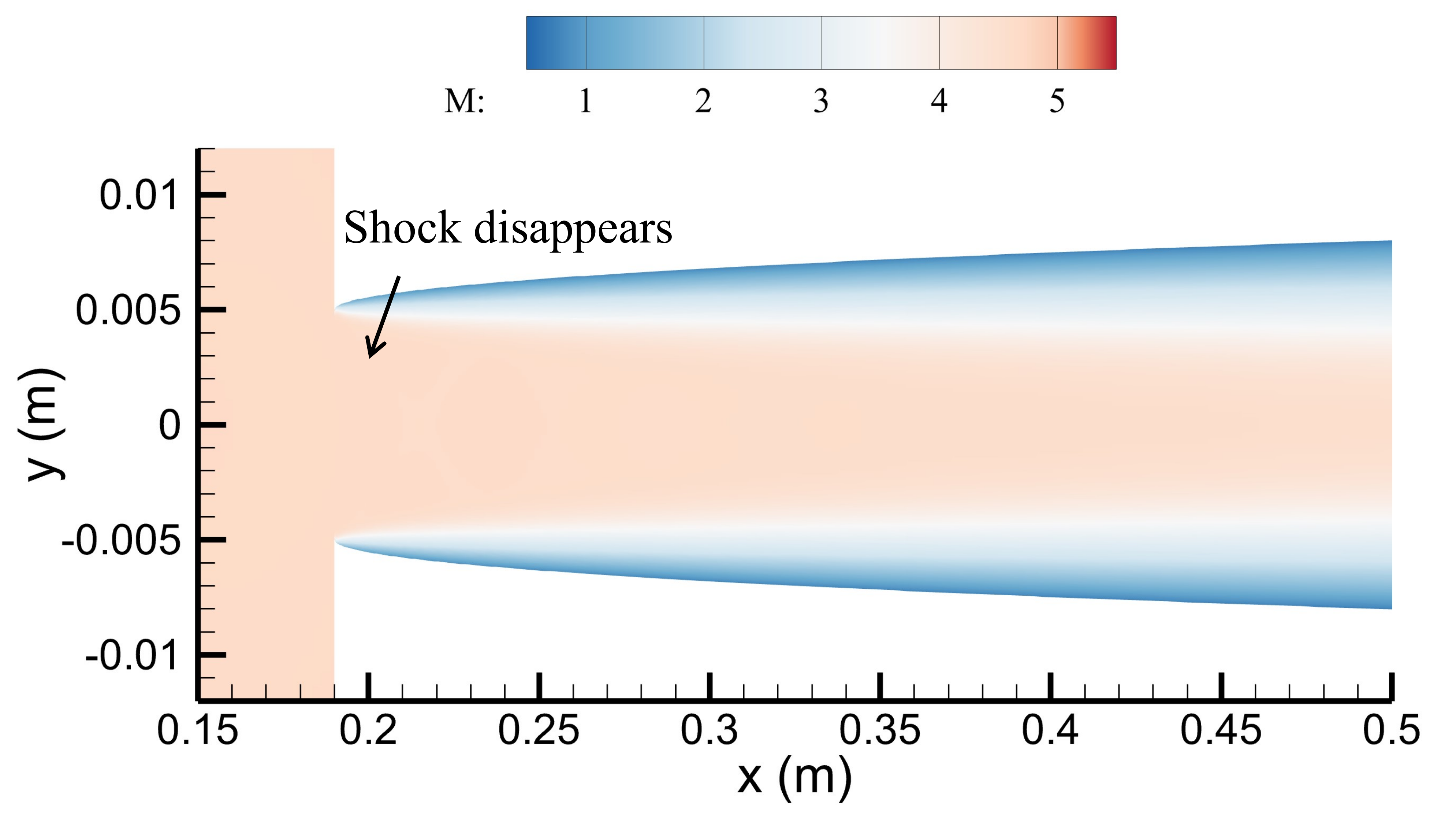}}
	\caption{\label{fig:machnd}Mach number distribution (a) before and (b) after changing the duct shape.}
\end{figure}

After calculating the frozen boundary layer equations, the displacement thickness distribution can be derived and expressed as $\delta _\mathrm{f}^{\ast}=Ix^{0.5}$, where I is a constant determined by the boundary layer profile and freestream conditions. Due to the absence of nonequilibrium effects in this calculation, the curve is scaled up proportionally. A modified duct with 1.2 times $\delta _\mathrm{f}^{\ast}$ effectively mitigates the leading-edge shock, as shown in Fig.~\ref{fig:machnd}. The duct with a constant area yields a prominent leading-edge shock, whereas the curved duct nearly eliminates it. A comparison of nonequilibrium temperatures in Fig.~\ref{fig:Temperaturepa} reveals that while there are minor fluctuations at the front, the curved duct yields a temperature profile closer to the inviscid result, thus significantly expanding the undisturbed zone.

\subsection{\label{ExmDetEMZ}Determination of the effective measurement zone}
The design of the shock tube conditions employ the one-dimensional Euler equations with a thermally perfect gas model and a ten-species gas model encompassing detonation products and test gas (H\textsubscript{2}, O\textsubscript{2}, H, O, OH, H\textsubscript{2}O, HO\textsubscript{2}, N\textsubscript{2}, N, NO)\cite{Luo2020}. It is an in-house code developed by the Institute of Mechanics, Chinese Academy of Sciences. The simulated shock velocity is required to be 15\% higher than that specified in Table~\ref{tab:shockWC} to account for shock wave attenuation. A shock velocity of 8.0 km/s is achieved with driver gas H\textsubscript{2}:O\textsubscript{2}:N\textsubscript{2} = 2:1:0.5 and \textit{P}\textsubscript{4} = 15 bar, which is sufficiently strong to induce a large amount of nitrogen dissociation and electronic excitation.

\begin{figure}
	\centering
	\includegraphics[width=0.45\textwidth]{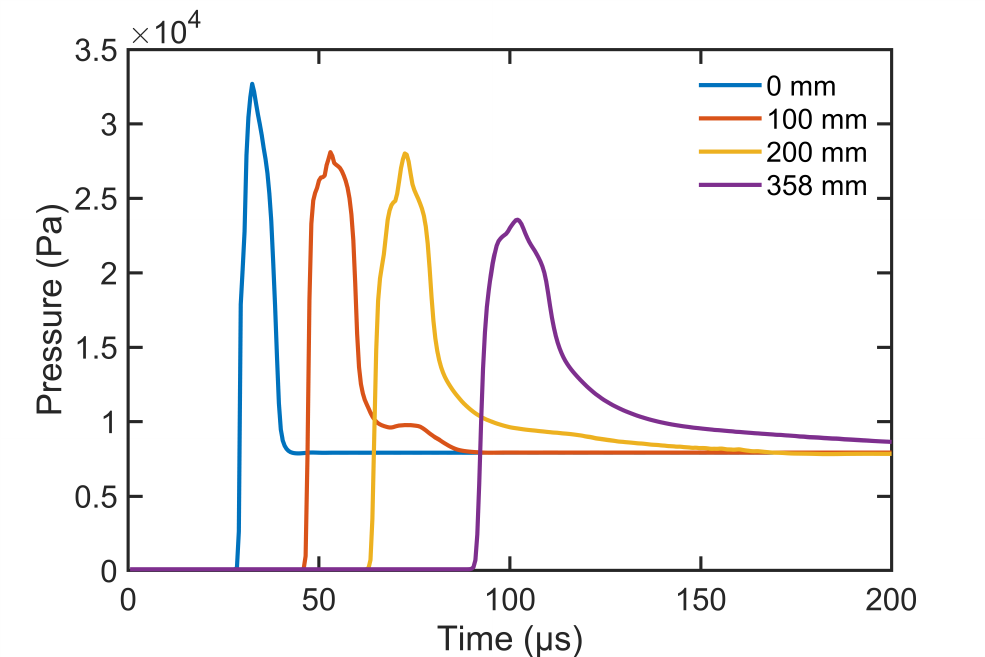}
	\caption{\label{fig:pressureha}Pressure histories at different data-probe locations within the observation window with the curved duct.}
\end{figure}

The computational domain, depicted in Fig.~\ref{fig:GridsB}, was employed for flow establishment simulations. To simplify the setup, the airfoil was extended to the duct front. Though this may not be practical in experiments, considering the hypersonic flow field, its impact on centerline flow establishment is minimal. The pressure histories at four centerline locations (0, 100, 200, and 358 mm from the duct's leading edge, with the latter corresponding to the diameter of the JF-10 shock tunnel's observation window) were examined. The moment when the head of the test gas exits the shock tube is considered the zero-point time. The pressure histories at different data-probe locations is shown in Fig.~\ref{fig:pressureha}. It is observed that the flow establishment time varied significantly, with probes farther from the leading edge needing more time. The shock tube test time can be predicted by solving the one-dimensional Euler equations or applying Mirels’ theory. They are tabulated with the flow establishment time simulated by Eilmer4 in Table~\ref{tab:effectivett}. \textit{Transit time} represents the time required for the gas to travel from the shock tube exit to the probe location. The end of the test time was calculated by adding transit time to the shock tube test time. \textit{Effective test time} denotes the time during which the local pressure was stable, calculated as the difference between the end and start of the test time, with negative values indicating no established flow at the corresponding location within the test time. The results align with Slack's theoretical analysis, showing distinct start, end, and effective test times at different locations, with the effective test time decreasing as the probe location moves away from the leading edge.

The Mirels’ theory predicted a shorter test time due to the consideration of viscous effects. To optimize OES measurements, the ideal scenario is to find an exposure time when each location of interest is within its effective test time. However, the duration of effective test time is staggered inside the duct. Beyond 100 mm from the leading edge, no effective test time is observed. Consequently, the measurement zone within the duct is approximately 100 mm under condition 4.
\begin{table*}
		 \renewcommand{\arraystretch}{1.2}
	\caption{\label{tab:effectivett}Effective test time at different locations with Mirels’ theory and inviscid simulation under condition 4.}
	\begin{ruledtabular}
		\begin{tabular}{lccccc}
			Methods & &0 mm &100 mm &200 mm &358 mm
			\\\hline
			\multirow{2}{*}{\makecell{Eilmer4 2-D viscous simulation}} 
			&Transit time ($\mu$s)    
			&21  &36  &51  &75
			\\
			&Start of test time ($\mu$s) &42  &85  &150  &249
			\\\hline
			\multirow{3}{*}{Mirels' theory}
			&Shock tube test time ($\mu$s) &\multicolumn{4}{c}{54}
			\\
			&End of test time ($\mu$s)    &75  &90  &105  &119
			\\
			&Effective test time ($\mu$s) &33  &5  &0  &0
			\\\hline
			\multirow{3}{*}{Inviscid shock tube simulation}
			&Shock tube test time ($\mu$s) &\multicolumn{4}{c}{66}
			\\
			&End of test time ($\mu$s)    &87  &102  &117  &141
			\\
			&Effective test time ($\mu$s) &56  &29  &0  &0
			\\
		\end{tabular}
	\end{ruledtabular}
\end{table*}

\begin{figure}
	\centering
	\includegraphics[width=0.45\textwidth]{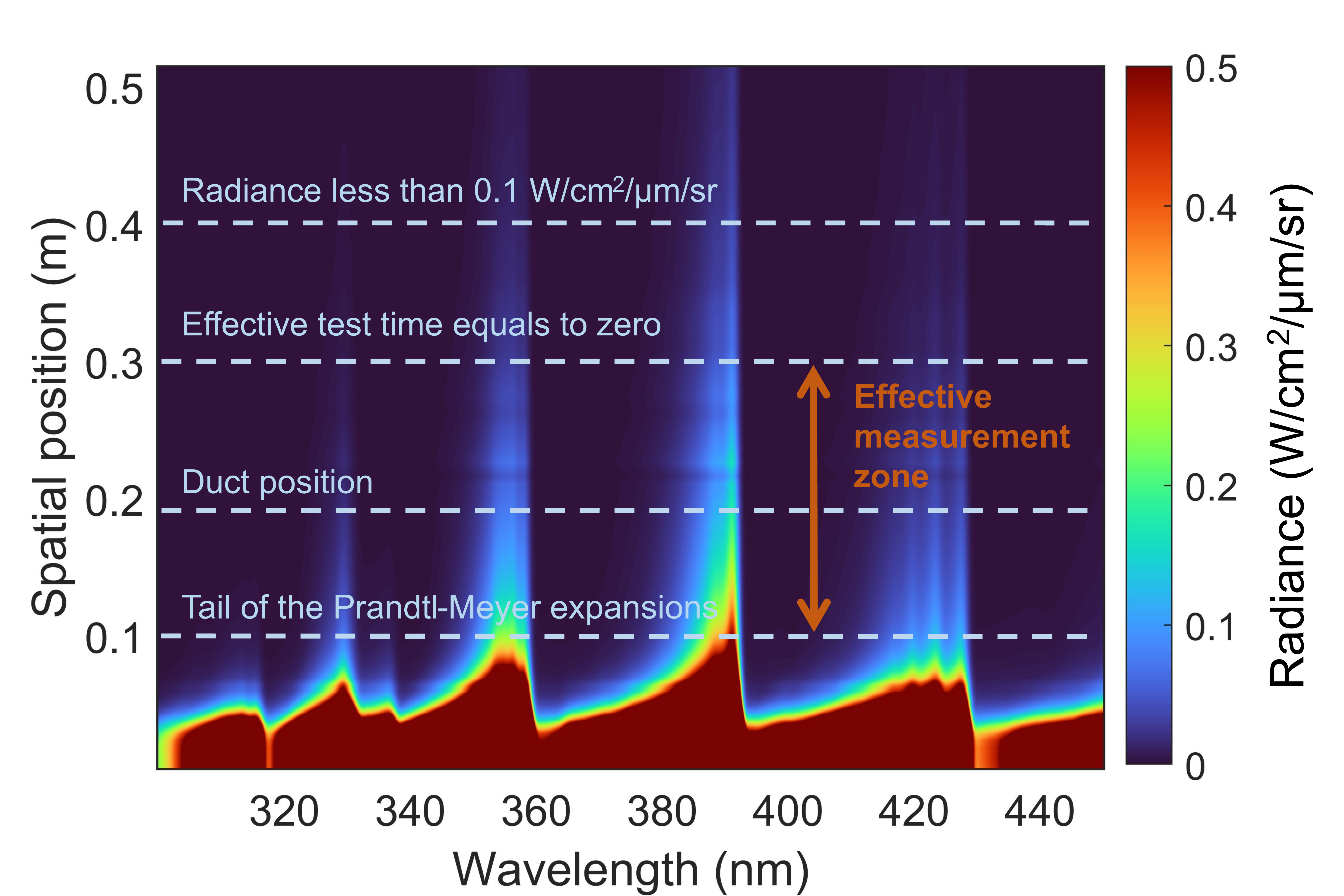}
	\caption{\label{fig:spetcralra}Spectral radiance along the centerline.}
\end{figure}

The unit Reynolds number at the duct inlet is $3\times10^5\;\mathrm{m^{-1}}$, which is much lower than the transition Reynolds number\cite{Chen1989,he1994}. Therefore, laminar boudary layer would be valid at the front position of the duct and transition will not significantly impact the measurement under this condition. With the employment of OES, the radiance on the centerline can be measured directly and the predicted three-dimensional spectral distribution is plotted in Fig.~\ref{fig:spetcralra}. Along the centerline, the radiance decreases due to the decrease in vibrational-electronic-electron temperature (\textit{T}\textsubscript{vee}), as shown in Fig.~\ref{fig:Temperaturepa}. Beyond 20 mm (\textit{x} = 40 mm) from the duct's leading edge, the radiance became lower than 0.1 W/cm\textsuperscript{2}/$\mu$m/sr, thus constraining the effective measurement zone. Generally, the length of the measurement zone under this condition is primarily limited by the effective test time, and it is around 200 mm after the duct is positioned.

\section{\label{Conclusions}conclusions}
An experimental configuration to study high-enthalpy radiating flows under nonequilibrium de-excitation is presented. A general design method is introduced, which combines theoretical analysis and numerical simulations. In general, this approach can be applied to various shock tube conditions, provided that equilibrium can be achieved at the shock tube outlet and appropriate diagnostic techniques can be employed. Using this method, different flow conditions can be achieved by adjusting the shock tube conditions and airfoil geometry, enabling the study of distinct behaviors of high-enthalpy radiating flows.

\textcolor{black}{The original design proposed by Wilson was unable to generate a sufficiently large undisturbed zone or a uniform flow along the centerline. With the aid of modern CFD techniques, significant modifications to the airfoil and duct expanded the undisturbed zone and achieved a uniform centerline flow, facilitating the measurement of nonequilibrium de-excitation. The pre-corner portion of Wilson's airfoil can be removed using the modified airfoil under most conditions. Additionally, the duct shape can be adjusted based on displacement thickness to eliminate the leading-edge shock. A qualitative analysis of the airfoil reveals that the undisturbed zone is determined by $\alpha$ and $D_\mathrm{af}$, with its maximum size constrained by the reflection of the Prandtl-Meyer expansion wave.} 

Based on the proposed method, an example of high-enthalpy N\textsubscript{2} flow with the employment of the JF-10 shock tube is detailed. A shock tube condition with \textit{V}\textsubscript{s} = 7 km/s and \textit{P}\textsubscript{1} = 100 Pa is appropriate for investigating the vibrational and electronic de-excitation under chemical freezing. The airfoil configuration is $\alpha=10^{\circ}$, $\beta=10^{\circ}$, $D_\mathrm{af}=22$ mm, and $L_\mathrm{af}>100$ mm. The implementation of a curved duct effectively mitigated the leading-edge shock and prolonged the undisturbed zone. The primary limitation of the effective measurement zone under this condition is the shock tube test time, since it is significantly limited under high-enthalpy conditions. This condition enables an undisturbed zone of approximately 200 mm in length. Therefore, this method achieves a comprehensive design of PMD and promotes further research on nonequilibrium de-excitation under high-enthalpy radiating flows.

\begin{acknowledgments}
This work was supported by the Hong Kong Research Grants Council (No. 15206522), the National Natural Science Foundation of China (No. 12202373 and 12232018), and the Strategic Priority Research Program of Chinese Academy of Sciences (No. XDB 0620203)..
\end{acknowledgments}

\section*{Data Availability Statement}

The data that support the findings of this study are available from the corresponding author upon reasonable request.

\bibliography{aipsamp}

\begin{thebibliography}{44}%
\makeatletter
\providecommand \@ifxundefined [1]{%
 \@ifx{#1\undefined}
}%
\providecommand \@ifnum [1]{%
 \ifnum #1\expandafter \@firstoftwo
 \else \expandafter \@secondoftwo
 \fi
}%
\providecommand \@ifx [1]{%
 \ifx #1\expandafter \@firstoftwo
 \else \expandafter \@secondoftwo
 \fi
}%
\providecommand \natexlab [1]{#1}%
\providecommand \enquote  [1]{``#1''}%
\providecommand \bibnamefont  [1]{#1}%
\providecommand \bibfnamefont [1]{#1}%
\providecommand \citenamefont [1]{#1}%
\providecommand \href@noop [0]{\@secondoftwo}%
\providecommand \href [0]{\begingroup \@sanitize@url \@href}%
\providecommand \@href[1]{\@@startlink{#1}\@@href}%
\providecommand \@@href[1]{\endgroup#1\@@endlink}%
\providecommand \@sanitize@url [0]{\catcode `\\12\catcode `\$12\catcode
  `\&12\catcode `\#12\catcode `\^12\catcode `\_12\catcode `\%12\relax}%
\providecommand \@@startlink[1]{}%
\providecommand \@@endlink[0]{}%
\providecommand \url  [0]{\begingroup\@sanitize@url \@url }%
\providecommand \@url [1]{\endgroup\@href {#1}{\urlprefix }}%
\providecommand \urlprefix  [0]{URL }%
\providecommand \Eprint [0]{\href }%
\providecommand \doibase [0]{http://dx.doi.org/}%
\providecommand \selectlanguage [0]{\@gobble}%
\providecommand \bibinfo  [0]{\@secondoftwo}%
\providecommand \bibfield  [0]{\@secondoftwo}%
\providecommand \translation [1]{[#1]}%
\providecommand \BibitemOpen [0]{}%
\providecommand \bibitemStop [0]{}%
\providecommand \bibitemNoStop [0]{.\EOS\space}%
\providecommand \EOS [0]{\spacefactor3000\relax}%
\providecommand \BibitemShut  [1]{\csname bibitem#1\endcsname}%
\let\auto@bib@innerbib\@empty
\bibitem [{\citenamefont {Reynier}(2016)}]{Reynier2016}%
  \BibitemOpen
  \bibfield  {author} {\bibinfo {author} {\bibfnamefont {P.}~\bibnamefont
  {Reynier}},\ }\bibfield  {title} {\enquote {\bibinfo {title} {Survey of
  high-enthalpy shock facilities in the perspective of radiation and chemical
  kinetics investigations},}\ }\href {\doibase 10.1016/j.paerosci.2016.04.002}
  {\bibfield  {journal} {\bibinfo  {journal} {Progress in Aerospace Sciences}\
  }\textbf {\bibinfo {volume} {85}},\ \bibinfo {pages} {1--32} (\bibinfo {year}
  {2016})}\BibitemShut {NoStop}%
\bibitem [{\citenamefont {Surzhikov}()}]{Surzhikov2007}%
  \BibitemOpen
  \bibfield  {author} {\bibinfo {author} {\bibfnamefont {S.}~\bibnamefont
  {Surzhikov}},\ }\bibfield  {title} {\enquote {\bibinfo {title} {Radiative gas
  dynamics of msl at angle of attack},}\ }in\ \href {\doibase
  10.2514/6.2016-0742} {\emph {\bibinfo {booktitle} {54th AIAA Aerospace
  Sciences Meeting}}}\BibitemShut {NoStop}%
\bibitem [{\citenamefont {Ray}\ \emph {et~al.}(2020)\citenamefont {Ray},
  \citenamefont {Kieweg}, \citenamefont {Dinzl}, \citenamefont {Carnes},
  \citenamefont {Weirs}, \citenamefont {Freno}, \citenamefont {Howard},
  \citenamefont {Smith}, \citenamefont {Nompelis},\ and\ \citenamefont
  {Candler}}]{Ray2020}%
  \BibitemOpen
  \bibfield  {author} {\bibinfo {author} {\bibfnamefont {J.}~\bibnamefont
  {Ray}}, \bibinfo {author} {\bibfnamefont {S.}~\bibnamefont {Kieweg}},
  \bibinfo {author} {\bibfnamefont {D.}~\bibnamefont {Dinzl}}, \bibinfo
  {author} {\bibfnamefont {B.}~\bibnamefont {Carnes}}, \bibinfo {author}
  {\bibfnamefont {V.~G.}\ \bibnamefont {Weirs}}, \bibinfo {author}
  {\bibfnamefont {B.}~\bibnamefont {Freno}}, \bibinfo {author} {\bibfnamefont
  {M.}~\bibnamefont {Howard}}, \bibinfo {author} {\bibfnamefont
  {T.}~\bibnamefont {Smith}}, \bibinfo {author} {\bibfnamefont
  {I.}~\bibnamefont {Nompelis}}, \ and\ \bibinfo {author} {\bibfnamefont
  {G.~V.}\ \bibnamefont {Candler}},\ }\bibfield  {title} {\enquote {\bibinfo
  {title} {Estimation of inflow uncertainties in laminar hypersonic double-cone
  experiments},}\ }\href {\doibase 10.2514/1.J059033} {\bibfield  {journal}
  {\bibinfo  {journal} {AIAA Journal}\ }\textbf {\bibinfo {volume} {58}},\
  \bibinfo {pages} {4461--4474} (\bibinfo {year} {2020})}\BibitemShut {NoStop}%
\bibitem [{\citenamefont {MacLean}\ and\ \citenamefont
  {Holden}(2006)}]{MacLean2006}%
  \BibitemOpen
  \bibfield  {author} {\bibinfo {author} {\bibfnamefont {M.}~\bibnamefont
  {MacLean}}\ and\ \bibinfo {author} {\bibfnamefont {M.}~\bibnamefont
  {Holden}},\ }\bibfield  {title} {\enquote {\bibinfo {title} {Numerical
  assessment of data in catalytic and transitional flows for martian entry},}\
  }in\ \href {\doibase 10.2514/6.2006-2946} {\emph {\bibinfo {booktitle} {9th
  AIAA/ASME Joint Thermophysics and Heat Transfer Conference}}}\ (\bibinfo
  {year} {2006})\BibitemShut {NoStop}%
\bibitem [{\citenamefont {Sharma}\ \emph
  {et~al.}(1993{\natexlab{a}})\citenamefont {Sharma}, \citenamefont {Ruffin},
  \citenamefont {Meyer}, \citenamefont {Gillespie},\ and\ \citenamefont
  {Yates}}]{Sharma1993}%
  \BibitemOpen
  \bibfield  {author} {\bibinfo {author} {\bibfnamefont {S.~P.}\ \bibnamefont
  {Sharma}}, \bibinfo {author} {\bibfnamefont {S.~M.}\ \bibnamefont {Ruffin}},
  \bibinfo {author} {\bibfnamefont {S.~A.}\ \bibnamefont {Meyer}}, \bibinfo
  {author} {\bibfnamefont {W.~D.}\ \bibnamefont {Gillespie}}, \ and\ \bibinfo
  {author} {\bibfnamefont {L.~A.}\ \bibnamefont {Yates}},\ }\bibfield  {title}
  {\enquote {\bibinfo {title} {Density measurements in an expanding flow using
  holographic interferometry},}\ }\href {\doibase 10.2514/3.415} {\bibfield
  {journal} {\bibinfo  {journal} {Journal of Thermophysics and Heat Transfer}\
  }\textbf {\bibinfo {volume} {7}},\ \bibinfo {pages} {261--268} (\bibinfo
  {year} {1993}{\natexlab{a}})}\BibitemShut {NoStop}%
\bibitem [{\citenamefont {McLaren}\ and\ \citenamefont
  {Appleton}(1970)}]{McLaren1970}%
  \BibitemOpen
  \bibfield  {author} {\bibinfo {author} {\bibfnamefont {T.~I.}\ \bibnamefont
  {McLaren}}\ and\ \bibinfo {author} {\bibfnamefont {J.~P.}\ \bibnamefont
  {Appleton}},\ }\bibfield  {title} {\enquote {\bibinfo {title} {Vibrational
  relaxation measurements of carbon monoxide in a shock‐tube expansion
  wave},}\ }\href {\doibase 10.1063/1.1674411} {\bibfield  {journal} {\bibinfo
  {journal} {The Journal of Chemical Physics}\ }\textbf {\bibinfo {volume}
  {53}},\ \bibinfo {pages} {2850--2857} (\bibinfo {year} {1970})}\BibitemShut
  {NoStop}%
\bibitem [{\citenamefont {Russo}(1967)}]{Russo1967}%
  \BibitemOpen
  \bibfield  {author} {\bibinfo {author} {\bibfnamefont {A.~L.}\ \bibnamefont
  {Russo}},\ }\bibfield  {title} {\enquote {\bibinfo {title}
  {Spectrophotometric measurements of the vibrational relaxation of co in
  shock‐wave and nozzle expansion‐flow environments},}\ }\href {\doibase
  10.1063/1.1701780} {\bibfield  {journal} {\bibinfo  {journal} {The Journal of
  Chemical Physics}\ }\textbf {\bibinfo {volume} {47}},\ \bibinfo {pages}
  {5201--5210} (\bibinfo {year} {1967})}\BibitemShut {NoStop}%
\bibitem [{\citenamefont {Ibraguimova}\ and\ \citenamefont
  {Shatalov}(2012)}]{Ibraguimova2012}%
  \BibitemOpen
  \bibfield  {author} {\bibinfo {author} {\bibfnamefont {L.}~\bibnamefont
  {Ibraguimova}}\ and\ \bibinfo {author} {\bibfnamefont {O.}~\bibnamefont
  {Shatalov}},\ }\enquote {\bibinfo {title} {Non-equilibrium kinetics behind
  shock waves experimental aspects},}\ in\ \href@noop {} {\emph {\bibinfo
  {booktitle} {High Temperature Phenomena in Shock Waves}}}\ (\bibinfo
  {publisher} {Springer},\ \bibinfo {year} {2012})\ pp.\ \bibinfo {pages}
  {99--147}\BibitemShut {NoStop}%
\bibitem [{\citenamefont {Blom}\ and\ \citenamefont {Pratt}(1968)}]{Blom1968}%
  \BibitemOpen
  \bibfield  {author} {\bibinfo {author} {\bibfnamefont {A.~P.}\ \bibnamefont
  {Blom}}\ and\ \bibinfo {author} {\bibfnamefont {N.~H.}\ \bibnamefont
  {Pratt}},\ }\bibfield  {title} {\enquote {\bibinfo {title} {Interpretation of
  band-reversal temperature measurements},}\ }\href {\doibase 10.2514/3.4700}
  {\bibfield  {journal} {\bibinfo  {journal} {AIAA Journal}\ }\textbf {\bibinfo
  {volume} {6}},\ \bibinfo {pages} {1172--1174} (\bibinfo {year}
  {1968})}\BibitemShut {NoStop}%
\bibitem [{\citenamefont {Brandis}\ \emph {et~al.}(2020)\citenamefont
  {Brandis}, \citenamefont {Saunders}, \citenamefont {Johnston}, \citenamefont
  {Cruden},\ and\ \citenamefont {White}}]{Brandis2020}%
  \BibitemOpen
  \bibfield  {author} {\bibinfo {author} {\bibfnamefont {A.~M.}\ \bibnamefont
  {Brandis}}, \bibinfo {author} {\bibfnamefont {D.~A.}\ \bibnamefont
  {Saunders}}, \bibinfo {author} {\bibfnamefont {C.~O.}\ \bibnamefont
  {Johnston}}, \bibinfo {author} {\bibfnamefont {B.~A.}\ \bibnamefont
  {Cruden}}, \ and\ \bibinfo {author} {\bibfnamefont {T.~R.}\ \bibnamefont
  {White}},\ }\bibfield  {title} {\enquote {\bibinfo {title} {Radiative heating
  on the after-body of martian entry vehicles},}\ }\href {\doibase
  10.2514/1.T5613} {\bibfield  {journal} {\bibinfo  {journal} {Journal of
  Thermophysics and Heat Transfer}\ }\textbf {\bibinfo {volume} {34}},\
  \bibinfo {pages} {66--77} (\bibinfo {year} {2020})}\BibitemShut {NoStop}%
\bibitem [{\citenamefont {Johnston}, \citenamefont {West},\ and\ \citenamefont
  {Brandis}(2019)}]{Johnston2019}%
  \BibitemOpen
  \bibfield  {author} {\bibinfo {author} {\bibfnamefont {C.~O.}\ \bibnamefont
  {Johnston}}, \bibinfo {author} {\bibfnamefont {T.~K.}\ \bibnamefont {West}},
  \ and\ \bibinfo {author} {\bibfnamefont {A.~M.}\ \bibnamefont {Brandis}},\
  }\bibfield  {title} {\enquote {\bibinfo {title} {Features of afterbody
  radiative heating for titan entry},}\ }in\ \href {\doibase
  10.2514/6.2019-3010} {\emph {\bibinfo {booktitle} {AIAA Aviation 2019
  Forum}}}\ (\bibinfo {year} {2019})\BibitemShut {NoStop}%
\bibitem [{\citenamefont {Johnston}\ and\ \citenamefont
  {Brandis}(2015)}]{Johnston2015}%
  \BibitemOpen
  \bibfield  {author} {\bibinfo {author} {\bibfnamefont {C.~O.}\ \bibnamefont
  {Johnston}}\ and\ \bibinfo {author} {\bibfnamefont {A.~M.}\ \bibnamefont
  {Brandis}},\ }\bibfield  {title} {\enquote {\bibinfo {title} {Features of
  afterbody radiative heating for earth entry},}\ }\href {\doibase
  10.2514/1.A33084} {\bibfield  {journal} {\bibinfo  {journal} {Journal of
  Spacecraft and Rockets}\ }\textbf {\bibinfo {volume} {52}},\ \bibinfo {pages}
  {105--119} (\bibinfo {year} {2015})}\BibitemShut {NoStop}%
\bibitem [{\citenamefont {Park}\ and\ \citenamefont {Lee}(1995)}]{Park1995}%
  \BibitemOpen
  \bibfield  {author} {\bibinfo {author} {\bibfnamefont {C.}~\bibnamefont
  {Park}}\ and\ \bibinfo {author} {\bibfnamefont {S.-H.}\ \bibnamefont {Lee}},\
  }\bibfield  {title} {\enquote {\bibinfo {title} {Validation of
  multitemperature nozzle flow code},}\ }\href {\doibase 10.2514/3.622}
  {\bibfield  {journal} {\bibinfo  {journal} {Journal of Thermophysics and Heat
  Transfer}\ }\textbf {\bibinfo {volume} {9}},\ \bibinfo {pages} {9--16}
  (\bibinfo {year} {1995})}\BibitemShut {NoStop}%
\bibitem [{\citenamefont {Gu}, \citenamefont {Hao},\ and\ \citenamefont
  {Wen}(2022)}]{Gu2022}%
  \BibitemOpen
  \bibfield  {author} {\bibinfo {author} {\bibfnamefont {S.}~\bibnamefont
  {Gu}}, \bibinfo {author} {\bibfnamefont {J.}~\bibnamefont {Hao}}, \ and\
  \bibinfo {author} {\bibfnamefont {C.-Y.}\ \bibnamefont {Wen}},\ }\bibfield
  {title} {\enquote {\bibinfo {title} {State-specific study of air in the
  expansion tunnel nozzle and test section},}\ }\href {\doibase
  10.2514/1.J061479} {\bibfield  {journal} {\bibinfo  {journal} {AIAA Journal}\
  }\textbf {\bibinfo {volume} {60}},\ \bibinfo {pages} {4024--4038} (\bibinfo
  {year} {2022})}\BibitemShut {NoStop}%
\bibitem [{\citenamefont {Gimelshein}\ and\ \citenamefont
  {Wysong}(2021)}]{Gimelshein2021}%
  \BibitemOpen
  \bibfield  {author} {\bibinfo {author} {\bibfnamefont {S.~F.}\ \bibnamefont
  {Gimelshein}}\ and\ \bibinfo {author} {\bibfnamefont {I.~J.}\ \bibnamefont
  {Wysong}},\ }\bibfield  {title} {\enquote {\bibinfo {title} {Nonequilibrium
  effects in high enthalpy gas flows expanding through nozzles},}\ }\href@noop
  {} {\bibfield  {journal} {\bibinfo  {journal} {Physics of Fluids}\ }\textbf
  {\bibinfo {volume} {33}} (\bibinfo {year} {2021})}\BibitemShut {NoStop}%
\bibitem [{\citenamefont {Holbeche}(1964)}]{Holbeche1964}%
  \BibitemOpen
  \bibfield  {author} {\bibinfo {author} {\bibfnamefont {T.~A.}\ \bibnamefont
  {Holbeche}},\ }\bibfield  {title} {\enquote {\bibinfo {title}
  {Spectrum-line-reversal temperature measurements through unsteady rarefaction
  waves in vibrationally relaxing oxygen},}\ }\href {\doibase 10.1038/203476b0}
  {\bibfield  {journal} {\bibinfo  {journal} {Nature}\ }\textbf {\bibinfo
  {volume} {203}},\ \bibinfo {pages} {476--479} (\bibinfo {year}
  {1964})}\BibitemShut {NoStop}%
\bibitem [{\citenamefont {Cleaver}\ and\ \citenamefont
  {Crow}(1973)}]{Cleaver1973}%
  \BibitemOpen
  \bibfield  {author} {\bibinfo {author} {\bibfnamefont {J.~W.}\ \bibnamefont
  {Cleaver}}\ and\ \bibinfo {author} {\bibfnamefont {I.~G.}\ \bibnamefont
  {Crow}},\ }\bibfield  {title} {\enquote {\bibinfo {title} {Vibrational
  relaxation of oxygen in an unsteady expansion wave},}\ }\href {\doibase
  10.1063/1.1680237} {\bibfield  {journal} {\bibinfo  {journal} {The Journal of
  Chemical Physics}\ }\textbf {\bibinfo {volume} {59}},\ \bibinfo {pages}
  {1592--1598} (\bibinfo {year} {1973})}\BibitemShut {NoStop}%
\bibitem [{\citenamefont {Nasser}\ and\ \citenamefont
  {Cleaver}(1977)}]{Nasser1977}%
  \BibitemOpen
  \bibfield  {author} {\bibinfo {author} {\bibfnamefont {A.~E.~M.}\
  \bibnamefont {Nasser}}\ and\ \bibinfo {author} {\bibfnamefont {J.~W.}\
  \bibnamefont {Cleaver}},\ }\bibfield  {title} {\enquote {\bibinfo {title}
  {Vibrational relaxation of carbon monoxide in an unsteady expansion wave},}\
  }\href {\doibase 10.1016/0094-5765(77)90056-X} {\bibfield  {journal}
  {\bibinfo  {journal} {Acta Astronautica}\ }\textbf {\bibinfo {volume} {4}},\
  \bibinfo {pages} {357--373} (\bibinfo {year} {1977})}\BibitemShut {NoStop}%
\bibitem [{\citenamefont {Tibère-Inglesse}\ \emph {et~al.}(2022)\citenamefont
  {Tibère-Inglesse}, \citenamefont {Bensassi}, \citenamefont {Brandis},\ and\
  \citenamefont {Cruden}}]{Tibère-Inglesse2022}%
  \BibitemOpen
  \bibfield  {author} {\bibinfo {author} {\bibfnamefont {A.~C.}\ \bibnamefont
  {Tibère-Inglesse}}, \bibinfo {author} {\bibfnamefont {K.}~\bibnamefont
  {Bensassi}}, \bibinfo {author} {\bibfnamefont {A.~M.}\ \bibnamefont
  {Brandis}}, \ and\ \bibinfo {author} {\bibfnamefont {B.~A.}\ \bibnamefont
  {Cruden}},\ }\bibfield  {title} {\enquote {\bibinfo {title} {Shock tube
  radiation measurement in expanding air flows},}\ }in\ \href {\doibase
  10.2514/6.2022-0117} {\emph {\bibinfo {booktitle} {AIAA Scitech 2022
  Forum}}}\ (\bibinfo {year} {2022})\BibitemShut {NoStop}%
\bibitem [{\citenamefont {Sharma}\ \emph
  {et~al.}(1993{\natexlab{b}})\citenamefont {Sharma}, \citenamefont {Ruffin},
  \citenamefont {Gillespie},\ and\ \citenamefont {Meyer}}]{Sharma1993VR}%
  \BibitemOpen
  \bibfield  {author} {\bibinfo {author} {\bibfnamefont {S.~P.}\ \bibnamefont
  {Sharma}}, \bibinfo {author} {\bibfnamefont {S.~M.}\ \bibnamefont {Ruffin}},
  \bibinfo {author} {\bibfnamefont {W.~D.}\ \bibnamefont {Gillespie}}, \ and\
  \bibinfo {author} {\bibfnamefont {S.~A.}\ \bibnamefont {Meyer}},\ }\bibfield
  {title} {\enquote {\bibinfo {title} {Vibrational relaxation measurements in
  an expanding flow using spontaneous raman scattering},}\ }\href {\doibase
  10.2514/3.479} {\bibfield  {journal} {\bibinfo  {journal} {Journal of
  Thermophysics and Heat Transfer}\ }\textbf {\bibinfo {volume} {7}},\ \bibinfo
  {pages} {697--703} (\bibinfo {year} {1993}{\natexlab{b}})}\BibitemShut
  {NoStop}%
\bibitem [{\citenamefont {Sebacher}\ and\ \citenamefont
  {Guy}(1974)}]{Sebacher1974}%
  \BibitemOpen
  \bibfield  {author} {\bibinfo {author} {\bibfnamefont {D.~I.}\ \bibnamefont
  {Sebacher}}\ and\ \bibinfo {author} {\bibfnamefont {R.~W.}\ \bibnamefont
  {Guy}},\ }\href@noop {} {\enquote {\bibinfo {title} {Vibrational relaxation
  in expanding {N}\textsubscript{2} and air},}\ }\bibinfo {type} {Tech. Rep.}\
  \bibinfo {number} {NASA-TM-X-71988}\ (\bibinfo  {institution} {NASA},\
  \bibinfo {year} {1974})\BibitemShut {NoStop}%
\bibitem [{\citenamefont {Bender}, \citenamefont {Mitchner},\ and\
  \citenamefont {Kruger}(1978)}]{Bender1978}%
  \BibitemOpen
  \bibfield  {author} {\bibinfo {author} {\bibfnamefont {D.~J.}\ \bibnamefont
  {Bender}}, \bibinfo {author} {\bibfnamefont {M.}~\bibnamefont {Mitchner}}, \
  and\ \bibinfo {author} {\bibfnamefont {C.~H.}\ \bibnamefont {Kruger}},\
  }\bibfield  {title} {\enquote {\bibinfo {title} {Measurement of vibrational
  population distributions in a supersonic expansion of carbon monoxide},}\
  }\href {\doibase 10.1063/1.862345} {\bibfield  {journal} {\bibinfo  {journal}
  {Physics of Fluids}\ }\textbf {\bibinfo {volume} {21}},\ \bibinfo {pages}
  {1073--1085} (\bibinfo {year} {1978})}\BibitemShut {NoStop}%
\bibitem [{\citenamefont {Park}(1968)}]{Park1968}%
  \BibitemOpen
  \bibfield  {author} {\bibinfo {author} {\bibfnamefont {C.}~\bibnamefont
  {Park}},\ }\bibfield  {title} {\enquote {\bibinfo {title} {Measurement of
  ionic recombination rate of nitrogen},}\ }\href {\doibase 10.2514/3.4937}
  {\bibfield  {journal} {\bibinfo  {journal} {AIAA Journal}\ }\textbf {\bibinfo
  {volume} {6}},\ \bibinfo {pages} {2090--2094} (\bibinfo {year}
  {1968})}\BibitemShut {NoStop}%
\bibitem [{\citenamefont {Park}(1973)}]{Park1973}%
  \BibitemOpen
  \bibfield  {author} {\bibinfo {author} {\bibfnamefont {C.}~\bibnamefont
  {Park}},\ }\bibfield  {title} {\enquote {\bibinfo {title} {Comparison of
  electron and electronic temperatures in recombining nozzle flow of ionized
  nitrogen–hydrogen mixture. part 2. experiment},}\ }\href {\doibase
  10.1017/S0022377800007443} {\bibfield  {journal} {\bibinfo  {journal}
  {Journal of Plasma Physics}\ }\textbf {\bibinfo {volume} {9}},\ \bibinfo
  {pages} {217--234} (\bibinfo {year} {1973})}\BibitemShut {NoStop}%
\bibitem [{\citenamefont {Igra}(1975)}]{Igra1975}%
  \BibitemOpen
  \bibfield  {author} {\bibinfo {author} {\bibfnamefont {O.}~\bibnamefont
  {Igra}},\ }\bibfield  {title} {\enquote {\bibinfo {title} {Supersonic
  expansion on non-equilibrium plasmas},}\ }\href {\doibase
  10.1016/0376-0421(75)90018-4} {\bibfield  {journal} {\bibinfo  {journal}
  {Progress in Aerospace Sciences}\ }\textbf {\bibinfo {volume} {16}},\
  \bibinfo {pages} {299--366} (\bibinfo {year} {1975})}\BibitemShut {NoStop}%
\bibitem [{\citenamefont {Gu}\ \emph {et~al.}(2021)\citenamefont {Gu},
  \citenamefont {Morgan}, \citenamefont {McIntyre},\ and\ \citenamefont
  {Brandis}}]{Gu2021}%
  \BibitemOpen
  \bibfield  {author} {\bibinfo {author} {\bibfnamefont {S.}~\bibnamefont
  {Gu}}, \bibinfo {author} {\bibfnamefont {R.~G.}\ \bibnamefont {Morgan}},
  \bibinfo {author} {\bibfnamefont {T.~J.}\ \bibnamefont {McIntyre}}, \ and\
  \bibinfo {author} {\bibfnamefont {A.~M.}\ \bibnamefont {Brandis}},\
  }\bibfield  {title} {\enquote {\bibinfo {title} {An experimental study of co2
  thermochemical nonequilibrium},}\ }\href {\doibase 10.2514/1.J061037}
  {\bibfield  {journal} {\bibinfo  {journal} {AIAA Journal}\ ,\ \bibinfo
  {pages} {1--10}} (\bibinfo {year} {2021})}\BibitemShut {NoStop}%
\bibitem [{\citenamefont {Kelly}, \citenamefont {Gildfind},\ and\ \citenamefont
  {McIntyre}(2021)}]{Kelly2021}%
  \BibitemOpen
  \bibfield  {author} {\bibinfo {author} {\bibfnamefont {R.~M.}\ \bibnamefont
  {Kelly}}, \bibinfo {author} {\bibfnamefont {D.~E.}\ \bibnamefont {Gildfind}},
  \ and\ \bibinfo {author} {\bibfnamefont {T.~J.}\ \bibnamefont {McIntyre}},\
  }\bibfield  {title} {\enquote {\bibinfo {title} {Emission spectroscopy of
  ionizing superorbital expanding flow},}\ }\href {\doibase 10.2514/1.J059345}
  {\bibfield  {journal} {\bibinfo  {journal} {AIAA Journal}\ }\textbf {\bibinfo
  {volume} {59}},\ \bibinfo {pages} {3217--3227} (\bibinfo {year}
  {2021})}\BibitemShut {NoStop}%
\bibitem [{\citenamefont {Wei}\ \emph {et~al.}(2017)\citenamefont {Wei},
  \citenamefont {Morgan}, \citenamefont {McIntyre}, \citenamefont {Brandis},\
  and\ \citenamefont {Johnston}}]{Wei2017}%
  \BibitemOpen
  \bibfield  {author} {\bibinfo {author} {\bibfnamefont {H.}~\bibnamefont
  {Wei}}, \bibinfo {author} {\bibfnamefont {R.~G.}\ \bibnamefont {Morgan}},
  \bibinfo {author} {\bibfnamefont {T.}~\bibnamefont {McIntyre}}, \bibinfo
  {author} {\bibfnamefont {A.~M.}\ \bibnamefont {Brandis}}, \ and\ \bibinfo
  {author} {\bibfnamefont {C.~O.}\ \bibnamefont {Johnston}},\ }\bibfield
  {title} {\enquote {\bibinfo {title} {Experimental and numerical investigation
  of air radiation in superorbital expanding flow},}\ }in\ \href {\doibase
  10.2514/6.2017-4531} {\emph {\bibinfo {booktitle} {47th AIAA Thermophysics
  Conference}}}\ (\bibinfo {year} {2017})\BibitemShut {NoStop}%
\bibitem [{\citenamefont {Wilson}(1963)}]{Wilson1963}%
  \BibitemOpen
  \bibfield  {author} {\bibinfo {author} {\bibfnamefont {J.}~\bibnamefont
  {Wilson}},\ }\bibfield  {title} {\enquote {\bibinfo {title} {An experiment to
  measure the recombination rate of oxygen},}\ }\href {\doibase
  10.1017/S0022112063000410} {\bibfield  {journal} {\bibinfo  {journal}
  {Journal of Fluid Mechanics}\ }\textbf {\bibinfo {volume} {15}},\ \bibinfo
  {pages} {497--512} (\bibinfo {year} {1963})}\BibitemShut {NoStop}%
\bibitem [{\citenamefont {Blom}\ and\ \citenamefont {Pratt}(1969)}]{Blom1969}%
  \BibitemOpen
  \bibfield  {author} {\bibinfo {author} {\bibfnamefont {A.~P.}\ \bibnamefont
  {Blom}}\ and\ \bibinfo {author} {\bibfnamefont {N.~H.}\ \bibnamefont
  {Pratt}},\ }\bibfield  {title} {\enquote {\bibinfo {title} {Vibrational
  energy excitation and de-excitation of co},}\ }\href {\doibase
  10.1038/2231052a0} {\bibfield  {journal} {\bibinfo  {journal} {Nature}\
  }\textbf {\bibinfo {volume} {223}},\ \bibinfo {pages} {1052--1053} (\bibinfo
  {year} {1969})}\BibitemShut {NoStop}%
\bibitem [{\citenamefont {Gu}\ \emph {et~al.}(2024)\citenamefont {Gu},
  \citenamefont {Hao}, \citenamefont {Wen}, \citenamefont {Hong},\ and\
  \citenamefont {Wang}}]{Gu2024}%
  \BibitemOpen
  \bibfield  {author} {\bibinfo {author} {\bibfnamefont {S.}~\bibnamefont
  {Gu}}, \bibinfo {author} {\bibfnamefont {J.}~\bibnamefont {Hao}}, \bibinfo
  {author} {\bibfnamefont {C.-Y.}\ \bibnamefont {Wen}}, \bibinfo {author}
  {\bibfnamefont {Q.}~\bibnamefont {Hong}}, \ and\ \bibinfo {author}
  {\bibfnamefont {Q.}~\bibnamefont {Wang}},\ }\bibfield  {title} {\enquote
  {\bibinfo {title} {How much does thermal nonequilibrium influence the overall
  atomic recombination during de-excitation?}}\ }\href {\doibase
  10.1016/j.chemphys.2024.112220} {\bibfield  {journal} {\bibinfo  {journal}
  {Chemical Physics}\ }\textbf {\bibinfo {volume} {580}},\ \bibinfo {pages}
  {112220} (\bibinfo {year} {2024})}\BibitemShut {NoStop}%
\bibitem [{\citenamefont {Mirels}(1963)}]{Mirels1963}%
  \BibitemOpen
  \bibfield  {author} {\bibinfo {author} {\bibfnamefont {H.}~\bibnamefont
  {Mirels}},\ }\bibfield  {title} {\enquote {\bibinfo {title} {Test time in
  low‐pressure shock tubes},}\ }\href {\doibase 10.1063/1.1706887} {\bibfield
   {journal} {\bibinfo  {journal} {Physics of Fluids}\ }\textbf {\bibinfo
  {volume} {6}},\ \bibinfo {pages} {1201--1214} (\bibinfo {year}
  {1963})}\BibitemShut {NoStop}%
\bibitem [{\citenamefont {Slack}\ \emph {et~al.}(1969)\citenamefont {Slack},
  \citenamefont {Bray}, \citenamefont {East},\ and\ \citenamefont
  {Pratt}}]{Slack1969}%
  \BibitemOpen
  \bibfield  {author} {\bibinfo {author} {\bibfnamefont {M.~W.}\ \bibnamefont
  {Slack}}, \bibinfo {author} {\bibfnamefont {K.~N.~C.}\ \bibnamefont {Bray}},
  \bibinfo {author} {\bibfnamefont {R.~A.}\ \bibnamefont {East}}, \ and\
  \bibinfo {author} {\bibfnamefont {N.~H.}\ \bibnamefont {Pratt}},\ }\bibfield
  {title} {\enquote {\bibinfo {title} {Steady expansion of shock‐heated gases
  for recombination studies},}\ }\href {\doibase 10.1063/1.1692588} {\bibfield
  {journal} {\bibinfo  {journal} {Physics of Fluids}\ }\textbf {\bibinfo
  {volume} {12}},\ \bibinfo {pages} {I--113} (\bibinfo {year}
  {1969})}\BibitemShut {NoStop}%
\bibitem [{\citenamefont {da~Silva}(2021)}]{daSilva2021}%
  \BibitemOpen
  \bibfield  {author} {\bibinfo {author} {\bibfnamefont {M.~L.}\ \bibnamefont
  {da~Silva}},\ }\href@noop {} {\emph {\bibinfo {title} {Spark line-by-line v.
  3.0 user’s manual}}} (\bibinfo {year} {2021})\BibitemShut {NoStop}%
\bibitem [{\citenamefont {Anderson}(2006)}]{Anderson2006}%
  \BibitemOpen
  \bibfield  {author} {\bibinfo {author} {\bibfnamefont {J.~D.}\ \bibnamefont
  {Anderson}},\ }\href@noop {} {\emph {\bibinfo {title} {Hypersonic and High
  Temperature Gas Dynamics}}}\ (\bibinfo  {publisher} {American Institute of
  Aeronautics and Astronautics},\ \bibinfo {address} {Reston},\ \bibinfo {year}
  {2006})\BibitemShut {NoStop}%
\bibitem [{\citenamefont {Gibbons}\ \emph {et~al.}(2023)\citenamefont
  {Gibbons}, \citenamefont {Damm}, \citenamefont {Jacobs},\ and\ \citenamefont
  {Gollan}}]{Gibbons2023}%
  \BibitemOpen
  \bibfield  {author} {\bibinfo {author} {\bibfnamefont {N.~N.}\ \bibnamefont
  {Gibbons}}, \bibinfo {author} {\bibfnamefont {K.~A.}\ \bibnamefont {Damm}},
  \bibinfo {author} {\bibfnamefont {P.~A.}\ \bibnamefont {Jacobs}}, \ and\
  \bibinfo {author} {\bibfnamefont {R.~J.}\ \bibnamefont {Gollan}},\ }\bibfield
   {title} {\enquote {\bibinfo {title} {Eilmer: An open-source multi-physics
  hypersonic flow solver},}\ }\href {\doibase 10.1016/j.cpc.2022.108551}
  {\bibfield  {journal} {\bibinfo  {journal} {Computer Physics Communications}\
  }\textbf {\bibinfo {volume} {282}},\ \bibinfo {pages} {108551} (\bibinfo
  {year} {2023})}\BibitemShut {NoStop}%
\bibitem [{\citenamefont {Browne}\ \emph {et~al.}(2018)\citenamefont {Browne},
  \citenamefont {Ziegler}, \citenamefont {Bitter}, \citenamefont {Schmidt},
  \citenamefont {Lawson},\ and\ \citenamefont {Shepherd}}]{Browne2018}%
  \BibitemOpen
  \bibfield  {author} {\bibinfo {author} {\bibfnamefont {S.}~\bibnamefont
  {Browne}}, \bibinfo {author} {\bibfnamefont {J.}~\bibnamefont {Ziegler}},
  \bibinfo {author} {\bibfnamefont {N.}~\bibnamefont {Bitter}}, \bibinfo
  {author} {\bibfnamefont {B.}~\bibnamefont {Schmidt}}, \bibinfo {author}
  {\bibfnamefont {J.}~\bibnamefont {Lawson}}, \ and\ \bibinfo {author}
  {\bibfnamefont {J.}~\bibnamefont {Shepherd}},\ }\href@noop {} {\enquote
  {\bibinfo {title} {Sdtoolbox: Numerical tools for shock and detonation wave
  modeling},}\ }\bibinfo {type} {Tech. Rep.}\ \bibinfo {number} {GALCIT Report
  FM2018. Vol. 1}\ (\bibinfo  {institution} {Explosion Dynamics Laboratory},\
  \bibinfo {year} {2018})\BibitemShut {NoStop}%
\bibitem [{\citenamefont {Kim}\ and\ \citenamefont {Jo}(2021)}]{Kim2021}%
  \BibitemOpen
  \bibfield  {author} {\bibinfo {author} {\bibfnamefont {J.~G.}\ \bibnamefont
  {Kim}}\ and\ \bibinfo {author} {\bibfnamefont {S.~M.}\ \bibnamefont {Jo}},\
  }\bibfield  {title} {\enquote {\bibinfo {title} {Modification of
  chemical-kinetic parameters for 11-air species in re-entry flows},}\ }\href
  {\doibase 10.1016/j.ijheatmasstransfer.2021.120950} {\bibfield  {journal}
  {\bibinfo  {journal} {International Journal of Heat and Mass Transfer}\
  }\textbf {\bibinfo {volume} {169}},\ \bibinfo {pages} {120950} (\bibinfo
  {year} {2021})}\BibitemShut {NoStop}%
\bibitem [{\citenamefont {Anderson}(2011)}]{Anderson2011}%
  \BibitemOpen
  \bibfield  {author} {\bibinfo {author} {\bibfnamefont {J.}~\bibnamefont
  {Anderson}},\ }\href@noop {} {\emph {\bibinfo {title} {Fundamentals of
  Aerodynamics}}}\ (\bibinfo  {publisher} {McGraw hill education},\ \bibinfo
  {year} {2011})\BibitemShut {NoStop}%
\bibitem [{\citenamefont {Rich}\ and\ \citenamefont
  {Treanor}(1970)}]{Rich1970}%
  \BibitemOpen
  \bibfield  {author} {\bibinfo {author} {\bibfnamefont {J.~W.}\ \bibnamefont
  {Rich}}\ and\ \bibinfo {author} {\bibfnamefont {C.~E.}\ \bibnamefont
  {Treanor}},\ }\bibfield  {title} {\enquote {\bibinfo {title} {Vibrational
  relaxation in gas-dynamic flows},}\ }\href {\doibase
  10.1146/annurev.fl.02.010170.002035} {\bibfield  {journal} {\bibinfo
  {journal} {Annual Review of Fluid Mechanics}\ }\textbf {\bibinfo {volume}
  {2}},\ \bibinfo {pages} {355--396} (\bibinfo {year} {1970})}\BibitemShut
  {NoStop}%
\bibitem [{\citenamefont {Zhao}\ \emph {et~al.}(2005)\citenamefont {Zhao},
  \citenamefont {Jiang}, \citenamefont {Saito}, \citenamefont {Lin},
  \citenamefont {Yu},\ and\ \citenamefont {Takayama}}]{Zhao2005}%
  \BibitemOpen
  \bibfield  {author} {\bibinfo {author} {\bibfnamefont {W.}~\bibnamefont
  {Zhao}}, \bibinfo {author} {\bibfnamefont {Z.~L.}\ \bibnamefont {Jiang}},
  \bibinfo {author} {\bibfnamefont {T.}~\bibnamefont {Saito}}, \bibinfo
  {author} {\bibfnamefont {J.~M.}\ \bibnamefont {Lin}}, \bibinfo {author}
  {\bibfnamefont {H.~R.}\ \bibnamefont {Yu}}, \ and\ \bibinfo {author}
  {\bibfnamefont {K.}~\bibnamefont {Takayama}},\ }\bibfield  {title} {\enquote
  {\bibinfo {title} {Performance of a detonation driven shock tunnel},}\ }\href
  {\doibase 10.1007/s00193-004-0238-1} {\bibfield  {journal} {\bibinfo
  {journal} {Shock Waves}\ }\textbf {\bibinfo {volume} {14}},\ \bibinfo {pages}
  {53--59} (\bibinfo {year} {2005})}\BibitemShut {NoStop}%
\bibitem [{\citenamefont {Luo}\ \emph {et~al.}(2020)\citenamefont {Luo},
  \citenamefont {Wang}, \citenamefont {Li}, \citenamefont {Li},\ and\
  \citenamefont {Zhao}}]{Luo2020}%
  \BibitemOpen
  \bibfield  {author} {\bibinfo {author} {\bibfnamefont {K.}~\bibnamefont
  {Luo}}, \bibinfo {author} {\bibfnamefont {Q.}~\bibnamefont {Wang}}, \bibinfo
  {author} {\bibfnamefont {J.~W.}\ \bibnamefont {Li}}, \bibinfo {author}
  {\bibfnamefont {J.~P.}\ \bibnamefont {Li}}, \ and\ \bibinfo {author}
  {\bibfnamefont {W.}~\bibnamefont {Zhao}},\ }\bibfield  {title} {\enquote
  {\bibinfo {title} {Numerical modeling of a high-enthalpy shock tunnel driven
  by gaseous detonation},}\ }\href {\doibase 10.1016/j.ast.2020.105958}
  {\bibfield  {journal} {\bibinfo  {journal} {Aerospace Science and
  Technology}\ }\textbf {\bibinfo {volume} {104}},\ \bibinfo {pages} {105958}
  (\bibinfo {year} {2020})}\BibitemShut {NoStop}%
\bibitem [{\citenamefont {Chen}, \citenamefont {Malik},\ and\ \citenamefont
  {Beckwith}(1989)}]{Chen1989}%
  \BibitemOpen
  \bibfield  {author} {\bibinfo {author} {\bibfnamefont {F.-J.}\ \bibnamefont
  {Chen}}, \bibinfo {author} {\bibfnamefont {M.~R.}\ \bibnamefont {Malik}}, \
  and\ \bibinfo {author} {\bibfnamefont {I.~E.}\ \bibnamefont {Beckwith}},\
  }\bibfield  {title} {\enquote {\bibinfo {title} {Boundary-layer transition on
  a cone and flat plate at mach 3.5},}\ }\href {\doibase 10.2514/3.10166}
  {\bibfield  {journal} {\bibinfo  {journal} {AIAA Journal}\ }\textbf {\bibinfo
  {volume} {27}},\ \bibinfo {pages} {687--693} (\bibinfo {year}
  {1989})}\BibitemShut {NoStop}%
\bibitem [{\citenamefont {He}\ and\ \citenamefont {Morgan}(1994)}]{he1994}%
  \BibitemOpen
  \bibfield  {author} {\bibinfo {author} {\bibfnamefont {Y.}~\bibnamefont
  {He}}\ and\ \bibinfo {author} {\bibfnamefont {R.~G.}\ \bibnamefont
  {Morgan}},\ }\bibfield  {title} {\enquote {\bibinfo {title} {Transition of
  compressible high enthalpy boundary layer flow over a flat plate},}\ }\href
  {\doibase 10.1017/S0001924000050181} {\bibfield  {journal} {\bibinfo
  {journal} {The Aeronautical Journal}\ }\textbf {\bibinfo {volume} {98}},\
  \bibinfo {pages} {25–34} (\bibinfo {year} {1994})}\BibitemShut {NoStop}%
\end{thebibliography}%


\begin{thebibliography}{0}%
\makeatletter
\providecommand \@ifxundefined [1]{%
 \@ifx{#1\undefined}
}%
\providecommand \@ifnum [1]{%
 \ifnum #1\expandafter \@firstoftwo
 \else \expandafter \@secondoftwo
 \fi
}%
\providecommand \@ifx [1]{%
 \ifx #1\expandafter \@firstoftwo
 \else \expandafter \@secondoftwo
 \fi
}%
\providecommand \natexlab [1]{#1}%
\providecommand \enquote  [1]{``#1''}%
\providecommand \bibnamefont  [1]{#1}%
\providecommand \bibfnamefont [1]{#1}%
\providecommand \citenamefont [1]{#1}%
\providecommand \href@noop [0]{\@secondoftwo}%
\providecommand \href [0]{\begingroup \@sanitize@url \@href}%
\providecommand \@href[1]{\@@startlink{#1}\@@href}%
\providecommand \@@href[1]{\endgroup#1\@@endlink}%
\providecommand \@sanitize@url [0]{\catcode `\\12\catcode `\$12\catcode
  `\&12\catcode `\#12\catcode `\^12\catcode `\_12\catcode `\%12\relax}%
\providecommand \@@startlink[1]{}%
\providecommand \@@endlink[0]{}%
\providecommand \url  [0]{\begingroup\@sanitize@url \@url }%
\providecommand \@url [1]{\endgroup\@href {#1}{\urlprefix }}%
\providecommand \urlprefix  [0]{URL }%
\providecommand \Eprint [0]{\href }%
\providecommand \doibase [0]{http://dx.doi.org/}%
\providecommand \selectlanguage [0]{\@gobble}%
\providecommand \bibinfo  [0]{\@secondoftwo}%
\providecommand \bibfield  [0]{\@secondoftwo}%
\providecommand \translation [1]{[#1]}%
\providecommand \BibitemOpen [0]{}%
\providecommand \bibitemStop [0]{}%
\providecommand \bibitemNoStop [0]{.\EOS\space}%
\providecommand \EOS [0]{\spacefactor3000\relax}%
\providecommand \BibitemShut  [1]{\csname bibitem#1\endcsname}%
\let\auto@bib@innerbib\@empty
\end{thebibliography}%

\end{document}